\documentclass[12pt]{article}

\usepackage{amsmath,verbatim,amssymb,epsfig,color,mathrsfs,graphics}
\usepackage{color}
\usepackage{float}

\def\squarebox#1{\hbox to #1{\hfill\vbox to #1{\vfill}}}
\def\boxit#1{\vbox{\hrule\hbox{\vrule\kern6pt
          \vbox{\kern6pt#1\kern6pt}\kern6pt\vrule}\hrule}}

\newtheorem{theorem}{Theorem}

\newcommand{\bJ}{\mbox{\bf J}}

\newcommand{\bM}{\mbox{\bf M}}

\newcommand{\bR}{\mbox{\bf R}}

\newcommand{\bX}{\mbox{\bf X}}
\newcommand{\bx}{\mbox{\bf x}}
\newcommand{\bQ}{\mbox{\bf Q}}

\newcommand{\bb}{\mbox{\bf b}}

\newcommand{\bG}{\mbox{\bf G}}
\newcommand{\bB}{\mbox{\bf B}}
\newcommand{\bA}{\mbox{\bf A}}

\newcommand{\bI}{\mbox{\bf I}}

\newcommand{\bU}{\mbox{\bf U}}

\newcommand{\bS}{\mbox{\bf S}}

\newcommand{\bW}{\mbox{\bf W}}
\newcommand{\bg}{\mbox{\bf g}}
\newcommand{\bY}{\mbox{\bf Y}}

\newcommand{\bbeta}{\boldsymbol{\beta}}

\newcommand{\bGamma}{\boldsymbol{\Gamma}}
\newcommand{\bSigma}{\boldsymbol{\Sigma}}
\newcommand{\balpha}{\boldsymbol{\alpha}}
\newcommand{\bOmega}{\boldsymbol{\Omega}}

\newcommand{\bmu}{\boldsymbol{\mu}}

\newcommand{\bgamma}{\boldsymbol{\gamma}}

\newcommand{\bPhi}{\boldsymbol{\Phi}}

\newcommand{\bPsi}{\boldsymbol{\Psi}}

\newcommand{\bphi}{\boldsymbol{\phi}}
\newcommand{\blambda}{\boldsymbol{\lambda}}
\def\qed{\hfill$\Box$\medskip}
\begin{document}

\baselineskip=22pt
\begin{center}
{\Large \bf Estimation of marginal model with subgroup auxiliary information}
\end{center}

\begin{center}

Jie He\\
{\it School of Mathematics, Beijing Normal University,
Beijing 100875, P. R. China}\\
hejie@mail.bnu.edu.cn\\

\vspace{2mm}

{ Xiaogang Duan, Shumei Zhang and Hui Li}\\
{\it School of Statistics, Beijing Normal University,
Beijing 100875, P. R. China}\\
xgduan@bnu.edu.cn, zsm1963@bnu.edu.cn, li\_hui@bnu.edu.cn\\

\end{center}


\begin{center}
{\sc Summary}
\end{center}
\par
Marginal model is a popular instrument for studying longitudinal data and cluster data. This paper investigates the estimator of marginal model with subgroup auxiliary information. To marginal model, we propose a new type of auxiliary information, and combine them with the traditional estimating equations of the quadratic inference function (QIF) method based on the generalized method of moments (GMM). Thus obtaining a more efficient estimator.
The asymptotic normality and the test statistics of the proposed estimator are established. The theoretical result shows that the estimator with subgroup information is more efficient than the conventional QIF one. Simulation studies are carried out to examine the performance of the proposed method under finite sample. We apply the proposed method to a real data for illustration.

\par\vspace{10mm}

\noindent {\it Key words:}
Auxiliary information; Estimation efficiency; Generalized method of moments; Longitudinal data; Marginal model; Quadratic inference function.

\vspace{1in}
\section{Introduction}
\par
Longitudinal data or cluster data exists commonly in many fields, such as biomedical, economics and so on. For longitudinal data, the unknown correlation structure within different measurements of the same subject brings many troubles to the analysis of this type of data. If the within-subject correlation is ignored and all observations are treated independently, the inference result may be inaccurate. As an extension of the generalized linear models (Nelder and Wedderburn, 1972; McCullagh and Nelder, 1989), Liang and Zeger (1986) proposed a kind of marginal model, which just make model assumption on the conditional expectation and variance of each component of the response given the covariates without considering the correlation structure. And they suggested to use the generalized estimating equations (GEE) method to estimate the parameters involved in this model. The score type estimating equations of the GEE method were obtained based on the working correlation matrix, which refers to the assumed conditional correlation structure among different components of the response vector.

The working correlation matrix usually contains an unknown nuisance parameter set which also needs to be estimated during the estimating procedure of GEE method. When the working correlation matrix is misspecified, although the GEE estimator is still consistent, it may suffer from loss of efficiency. Qu et al. (2000) introduced a more efficient
estimator obtained by the quadratic inference functions (QIF) method. They approximated the inverse of the working correlation matrix as a linear combination of some known basis matrices and constructed a quadratic inference function based on those matrices. Minimizing this quadratic function, the optimal solution is the QIF estimator. 
Since the nuisance parameters in working correlation matrix are not included in the quadratic function, the QIF estimator  still performs well in efficiency under the misspecified case. QIF method has been widely used in many models. Qu and Li (2006) studied the estimation of varying coefficient model with the QIF method. Li and Yin (2009) applied the QIF method to the accelerated failure time model with multivariate failure time. Li et al. (2016) investigated the QIF estimator of the marginal additive hazards model with cluster failure time.

Recently, how to apply the information from other sources to improve the efficiency of statistical inference is becoming a research focus, especially for the combination of the individual-level data and the summary-level information which can be obtained from other studies. Auxiliary information method is one popular approach, which stands for the information with specific form  provided by other datasets. For example, using covariate-specific disease prevalence information, that is the conditional probability of disease prevalence under different levels of covariate, as auxiliary information, Qin et al. (2015) obtained more efficient estimator to Logistic regression model in case-control studies. Chatterjee et al. (2016) developed the auxiliary information to regression models. They calculated the efficient likelihood estimator of parameter in regression model by incorporating the summary-level information from external big data with the likelihood function and extended their method to the case that the distribution of covariates in the internal data is different from that of the external data. Huang et al. (2016) proposed a double empirical likelihood estimator of the regression parameter in Cox's proportional hazards model which synthesizes the $t^*$-year survival probabilities as auxiliary information. Compared with the conventional partial likelihood estimator, the efficiency of the double empirical likelihood estimator has been improved significantly with the subgroup information.

In this paper, to improve the efficiency of parameter estimator in the marginal model, we add a new type of auxiliary information into the procedure of parameter estimation based on the GMM (Hansen, 1982) method. Different from the previous researches about the auxiliary information in regression models, we construct the estimator with auxiliary information from the estimating equations 
rather than the likelihood function. In addition, previous studies are mainly focus on the one dimensional independent response case, and we explore the multivariate correlated case directly.

The rest of the article is organized as follows. In section 2, we introduce the main results in this paper, which includes the marginal model as well as its properties, the proposed auxiliary information and the estimation procedure with auxiliary information by the GMM method. The asymptotic properties of estimator based on the procedures is also presented. The simulation studies are shown in section 3. And we illustrate our proposed procedures with a real data example in section 4. A brief discussion is given in Section 5. And the proofs of the theorems are in the Appendix.

\section{Main Results}
\par
\subsection{Marginal Model and Auxiliary Information}

In this paper, we just consider the case of longitudinal data, while the estimation procedure of cluster data is similar. For $i=1,\cdots, n$ and $j=1,\cdots,q$, let $\bY_i=(Y_{i1},\cdots,Y_{iq})^T$ be the response vector of the $i$th subject, $\bX^T_{ij}=(X_{ij1},\cdots,X_{ijp})$ be the $j$th observation of the $p$-dimensional covariate of the $i$th subject, thus $\bX_i=(X_{ijk})$ represents a $q \times p$ covariate matrix of the $i$th subject. Without loss of generality, we assume that observations among different subjects are independent. The marginal model takes the form of
\begin{eqnarray*}\label{margin}
&&h(\mu_{ij})=\bX^T_{ij}\bbeta,\\
&&\nu_{ij}=\psi v\left(\mu_{ij}\right), i=1,\cdots,n;j=1,\cdots,q,
\notag
\end{eqnarray*}
where $\mu_{ij}={\mathrm E}(Y_{ij}|\bX_{ij}=\bx_{ij})$, $\nu_{ij}={\mathrm {Var}}(Y_{ij}|\bX_{ij}=\bx_{ij})$ and $\bbeta$ is the parameter vector of interest. In addition, $\psi$ is the scalar parameter, $h(\cdot)$ and $v(\cdot)$ are known link functions.

Qu et al. (2000) proposed to estimate $\bbeta$ by the QIF method. They expressed the inverse of the working correlation matrix $\bR(\balpha)$ as
\begin{equation*}\label{basis}
\bR^{-1}(\balpha)=\sum_{l=1}^L \alpha_l\bM_l,
\end{equation*}
where $\balpha=(\alpha_1,\cdots,\alpha_L)^T$ is the nuisance parameter vector and $\left\{\bM_1,\cdots,\bM_L\right\}$ is a set of known basis matrices. Based on those basis matrices, they obtained the following estimating equation
\begin{equation}\label{score1}
\bS_n(\bbeta)=\frac{1}{n}\sum_{i=1}^n\bS(\bbeta,\bX_i)=\frac{1}{n}\sum_{i=1}^n\begin{pmatrix} \dot{\bmu}_i^T\bA_i^{-\frac{1}{2}}\bM_1\bA_i^{-\frac{1}{2}}(\bY_i-\bmu_i) \\ \vdots\\
                                 \dot{\bmu}_i^T\bA_i^{-\frac{1}{2}}\bM_L\bA_i^{-\frac{1}{2}}(\bY_i-\bmu_i) \end{pmatrix}
=\begin{pmatrix} \bS_n^{(1)}(\bbeta) \\ \vdots \\\bS_n^{(L)}(\bbeta)\end{pmatrix},
\end{equation}
where $\dot{\bmu}_i$ is the partial derivative of $\bmu_i$ with respect to $\bbeta^T$, $\bmu_i=(\mu_{i1},\cdots,\mu_{iq})^T$ is the conditional mean vector and $\bA_i$ is a diagonal matrix with each entry as the marginal conditional variance, ${\mathrm {Var}}(Y_{ij}|\bX_{ij}=\bx_{ij})$. The QIF estimator of $\bbeta$ is calculated by minimizing the quadratic inference function
\begin{equation*}
Q_n^{\ast}(\bbeta)=\bS_n(\bbeta)^T\left\{\bSigma_n^{\ast}\left(\bbeta\right)\right\}^{-1}\bS_n\left(\bbeta\right),
\end{equation*}
where $\bSigma_n^{\ast}(\bbeta)=n^{-1}\sum\limits_{i=1}^n\bS(\bbeta,\bX_i)\bS(\bbeta,\bX_i)^T$.

To marginal model, we suggest a new type of auxiliary information. Let $(\Omega_1,\cdots,\Omega_K)$ be a partition of $\Omega$, which is the range space of covariate $\bX$. If the conditional expectation of the response in subgroups $\Omega_k, k=1,\cdots,K$ are provided, we could consider them as auxiliary information. In fact, the specific expression of the auxiliary information is
\begin{equation*}
{\rm E}(\bY|\bX \in \Omega_k)=\bphi_k, k=1,\cdots,K.
\end{equation*}
Now, we change the auxiliary information to the form of estimating equations. By double expectation, $\bphi_k$ satisfies
\begin{equation}\label{aux}
{\rm E}\left[I(\bX \in \Omega_k)\left\{{\rm E}(\bY|\bX)-\bphi_k\right\}\right]=\mathbf{0}, k=1,\cdots,K.
\end{equation}
Define $\bPsi_k(\bbeta,\bX)=I(\bX \in \Omega_k)\left\{{\rm E}(\bY|\bX)-\bphi_k\right\}$, (\ref{aux}) is equivalent to
\begin{equation}\label{auxequ}
{\rm E}\left\{\bPsi_k(\bbeta,\bX)\right\}=\mathbf{0}, k=1,\cdots,K.
\end{equation}
In the following part, we will incorporate equation (\ref{auxequ}) into the estimate procedure.

\subsection{GMM Estimator with Auxiliary Information}
Noting (\ref{auxequ}), we have
\begin{equation}\label{auxinf}
\dfrac{1}{n}\sum_{i=1}^n \bPsi_k(\bbeta,\bX_i)=\mathbf{0},k=1,\cdots,K.
\end{equation}
Combing the estimating equations (\ref{auxinf}) with (\ref{score1}), we have
\begin{eqnarray}\label{estmating}
\bg_n(\bbeta)&=&\frac{1}{n}\sum_{i=1}^n \bg(\bbeta,\bX_i)
\notag\\
&=&\frac{1}{n}\sum_{i=1}^n \begin{pmatrix} \dot{\bmu}_i^T\bA_i^{-\frac{1}{2}}\bM_1\bA_i^{-\frac{1}{2}}(\bY_i-\bmu_i) \\
                        \vdots\\
                                 \dot{\bmu}_i^T\bA_i^{-\frac{1}{2}}\bM_L\bA_i^{-\frac{1}{2}}(\bY_i-\bmu_i) \\
                        \bPsi_1(\bbeta,\bX_i)\\
                        \vdots\\
                         \bPsi_K(\bbeta,\bX_i)
                 \end{pmatrix}
=\begin{pmatrix} \bS_n^{(1)}(\bbeta) \\ \vdots \\ \bS_n^{(L)}(\bbeta)\\ \bPsi_n^{(1)}(\bbeta)\\\vdots \\ \bPsi_n^{(K)}(\bbeta) \end{pmatrix},
\end{eqnarray}
where
$\bPsi_n^{(k)}(\bbeta)=n^{-1}\sum\limits_{i=1}^n \bPsi_k(\bbeta,\bX_i), k=1,\cdots,K$.

It is obvious that the number of estimating equations in (\ref{estmating}) is $pL+Kq$, which is larger than the dimension of parameter vector $\bbeta$. As it is stated in Hansen (1982), instead of solving the estimating equations directly, we estimate $\bbeta$ by minimizing the following quadratic function
\begin{equation*}
Q_n(\bbeta)=\bg_n(\bbeta)^T\left\{\bSigma_n(\bbeta)\right\}^{-1}\bg_n(\bbeta),
\end{equation*}
where $\bSigma_n(\bbeta)=n^{-1}\sum\limits_{i=1}^n\bg(\bbeta,\bX_i)\bg(\bbeta,\bX_i)^T$. We can obtain the optimal solution of $\bbeta$ by the Newton-Raphson iterative algorithm.

\subsection{Large sample properties}

We present the large sample properties of the proposed estimation method. Throughout, ``$\stackrel{\cal{D}}{\longrightarrow}$" denotes convergence in distribution.
\begin{theorem}
\label{thm1}
Under Conditions C1--C5 in the Appendix, we have that
\begin{equation*}
n^{1/2}(\widehat{\bbeta}_{\rm GMM}-\bbeta_0)\stackrel{\cal{D}}{\longrightarrow}N_p
\left \{{\bf 0},\big (\bB_1^T \bSigma_1^{-1} \bB_1 + \bB_2^T \bSigma_2^{-1}\bB_2)^{-1} \right\},
\end{equation*}
where $\bbeta_0$ is the true value of parameter vector $\bbeta$, and the definition of $\bB_1$, $\bB_2$, $\bSigma_1$ and $\bSigma_2$ are presented in the Appendix.
\end{theorem}

From the proof of Theorem{\ref{thm1}}, we have that the asymptotic variance of the QIF estimator is $(\bB_1^T \bSigma_1^{-1} \bB_1)^{-1}$. Since
\begin{equation*}
(\bB_1^T \bSigma_1^{-1} \bB_1 + \bB_2^T \bSigma_2^{-1}\bB_2)^{-1}\leq (\bB_1^T \bSigma_1^{-1} \bB_1)^{-1},
\end{equation*}
the new estimator with auxiliary information is asymptotically more efficient than the QIF one.

In order to make the statistical inference on $\bbeta$ in the marginal model, we construct $\chi^2$ test statistic on the basis of the quadratic inference function. Suppose that the parameter vector could be decomposed as $\bbeta=(\bgamma^T,\blambda^T)^T$, where $\bgamma$ is $p_1$ dimensional and $\blambda$ is $p-p_1$ dimensional. Suppose that we are interested in the hypothesis test $H_0:\bgamma=\bgamma_0$, then this hypothesis could be performed based on the following result by treating  $\blambda$ as a nuisance parameter vector.

\begin{theorem}
\label{thm2}
Let $\widehat{\bbeta}_{\rm GMM}=(\widehat{\bgamma}^T, \widehat{\blambda}^T)^T$ and $\widetilde{\blambda}$ be the {\rm GMM} estimator of $\blambda$ with auxiliary information when $\bgamma$ is fixed at $\bgamma_0$. Under Conditions C1--C5 in the Appendix, we have $n\left\{Q_n(\bgamma_0,\widetilde{\blambda})-Q_n(\widehat{\bgamma},\widehat{\blambda})\right\} \stackrel{\cal{D}}{\longrightarrow}\chi_{p_1}^2$.
\end{theorem}

The proof of Theorems \ref{thm1} and \ref{thm2} are briefly outlined in the Appendix.


\section{Simulation Studies}
In this section, we conduct a series of simulation studies to examine the performance of our proposed method under finite sample. We consider the following marginal model,
\begin{eqnarray}\label{mod}
{\rm E}(Y_{ij}|X_{ij1}={\rm x}_{ij1},X_{i2}={\rm x}_{i2})=\beta_1{\rm x}_{ij1}+\beta_2{\rm x}_{i2}, i=1,\cdots,n; j=1,2,3,
\end{eqnarray}
where $\beta_1=0.5$ and $\beta_2=-0.5$. Covariate $\bX_{1}=(X_{11},X_{21},X_{31})^T$ is generated from a multivariate normal distribution $N\left(\mathbf{0},\bSigma_X^1\right)$, and covariate $X_{2}$ is simulated from a Bernoulli distribution taking a value of 0 or 1 with probability $0.5$. We obtain the response vector $\bY=(Y_1,Y_2,Y_3)^T$ from a multivariate normal distribution with mean vector $\bmu=(\mu_{1},\mu_{2},\mu_{3})^T$ and covariance matrix $\bSigma_Y$, where $\mu_{j}=\beta_1{\rm x}_{j1}+\beta_2{\rm x}_2,j=1,2,3$.

In each case, we estimate $\beta_1$ and $\beta_2$ by the QIF and GMM with auxiliary information method, respectively. All simulation results are based on $500$ replications, which include the bias (Bias), the standard deviation (SD), the standard error (SE) and the empirical coverage probability (CP).

In order to test the influence of the working correlation matrix on the proposed method, we consider two common types of $\bSigma_Y$ and the working correlation matrix, which contains the compound symmetry (CS) structure and the first-order autoregressive correlation (AR(1)) structure. To covariate $\bX_1$, we assume that $\bSigma_X^1$ has CS structure with $\rho_X=0.5$, i.e., $\bSigma_X^{1}=\bI_3+\rho_X\left({\bf 1}_3{\bf 1}_3^T-\bI_3\right)$, where $\bI_3$ and ${\bf 1}_3$ represent the $3\times 3$ identity matrix and $3$-vector of ones respectively. The inverse of the working correlation matrix with CS structure can be decomposed as $\bR^{-1}\left(\alpha\right)=a_0(\alpha)\bI_3+a_1(\alpha)\bM_1$, where $a_0(\alpha)=-\left(\alpha+1\right)/\left(4\alpha^2-\alpha-1\right)$ and $a_1(\alpha)=\alpha/\left(4\alpha^2-\alpha-1\right)$, $\alpha$ is a nuisance parameter and $\bM_1$ is a $3\times 3$ matrix with $0$ on diagonal as well as $1$ off diagonal. Under the AR(1) assumption, the inverse of working correlation can be approximately expressed as $\bR^{-1}\left(\alpha\right)=b_0(\alpha)\bI_3+b_1(\alpha)\bM_2$ by omitting an unimportant matrix with $1$ on $(0,0)$  and $(3,3)$, and $0$ elsewhere. In the above expression, $b_0(\alpha)=(1+\alpha^2)/(1-\alpha^2)$, $b_1(\alpha)=-\alpha/(1-\alpha^2)$ and $\bM_2$ is a $3\times 3$ matrix with $1$ on two main off-diagonals and $0$ elsewhere. We then divide all the subjects into two subgroups by the value of covariate $X_2$, which have the follow form,
\begin{eqnarray*}
&&\Omega_1^{*}=\left\{X_2:X_2=1\right\},\\
&&\Omega_2^{*}=\left\{X_2:X_2=0\right\}.
\end{eqnarray*}
By the property of multivariate normal distribution, $\mathrm{E}\left(\bY\mid X_2=x_2\right)$ $=\left(\beta_2x_2,\cdots,\beta_2x_2\right)^T$. Substituting $\beta_2$ by its true value, the auxiliary information of the two groups are $\bphi_1^{*}=\left(-0.5,-0.5,-0.5\right)^T$ and $\bphi_2^{*}=\left(0,0,0\right)^T$, respectively. Choosing the sample size $n=300$, $\rho_Y=0.2,0.5$ and $0.8$, the simulation results are presented in Table \ref{Tab-1}.

In Table \ref{Tab-1}, ``GMMAI" represents the estimator obtained by the GMM method with auxiliary information $\bphi_1^{*}$ and $\bphi_2^{*}$. The results show that both the QIF and GMM incorporated auxiliary information methods perform well: the biases are very small, the SDs are close to the SEs, which are achieved by the asymptotic variance formula, and the CPs generally match the nominal level $95\%$.
The QIF and GMMAI estimators are more efficient when the the working correlation matrix is correctly specified than misspecified. However, the difference is not significate. Furthermore, when incorporated the auxiliary information, the estimators of $\beta_1$ are nearly of the same with the QIF ones as $\bphi_1^{*}$ and $\bphi_2^{*}$ only involve the information about covariate $X_2$. Whereas, the results of $\beta_2$ by these two methods are quite different: the SDs of $\hat{\beta}_2$ by the GMM method with auxiliary information are only about $1/2$ to those by the QIF method in all cases, which shows that the efficiency of parameter estimation can be improved largely when considering the auxiliary information. Since whether the correlation matrix is specified correctly has little influence on the estimation results, we just consider the cases with correct specified $\bR(\alpha)$  
in the following simulation studies.

Now, we study the effect of auxiliary information on estimation efficiency in detail. 
We consider the values of $\bX_1$ as well as $X_2$ when grouping the subjects. The obtained subgroups can be summarized as
\begin{eqnarray*}
&&\Omega_1=\left\{\left(\bX_1,X_2\right)\mid X_{11}\geq 0,X_2=1\right\},\\
&&\Omega_2=\left\{\left(\bX_1,X_2\right)\mid X_{11}< 0,X_2=1\right\},\\
&&\Omega_3=\left\{\left(\bX_1,X_2\right)\mid X_{11}\geq 0,X_2=0\right\},\\
&&\Omega_4=\left\{\left(\bX_1,X_2\right)\mid X_{11}< 0,X_2=0\right\}.
\end{eqnarray*}
To estimate the auxiliary information, we calculate the mean of $\bY$ in each subgroup, and express the auxiliary information as $\bphi_1$, $\bphi_2$, $\bphi_3$ and $\bphi_4$. Once combined $\Omega_1$ and $\Omega_2$, $\Omega_3$ and $\Omega_4$ respectively, $\Omega_1-\Omega_4$ will shrink to $\Omega_1^{*}$ and $\Omega_2^{*}$.

Firstly, we consider a simple case when $\bSigma_X^1=\bI_3$, $\bSigma_Y$ has CS and AR(1) structure with $\rho_Y=0.2,0.5$ and $0.8$, sample size $n=200$ and $500$. We estimate the parameters in model (\ref{mod}) by three methods--QIF, GMMAI2 and GMMAI4, where ``GMMAI2" represents GMM estimator with auxiliary information $\bphi_1^{*}-\bphi_2^{*}$ and ``GMMAI4" stands for GMM estimator with subgroup information $\bphi_1-\bphi_4$. The simulation results are shown in Table \ref{Tab-2}. The results of GMMAI2 in Table \ref{Tab-2} are similar with that in Table \ref{Tab-1}, that is only the efficiency of $\beta_2$ can be improved when we incorporate the auxiliary information $\bphi_1^{*}-\bphi_2^{*}$. However, when we combine $\bphi_1-\bphi_4$ with the estimation procedure, the power of $\beta_1$ is also improved, at the same time, the SD of $\beta_2$ is more smaller than GMMAI2 method. For example, when $n=200$ and $\bSigma_Y$ has CS structure with $\rho_Y=0.2$, the SD of  ${\hat\beta}_1$ by GMMAI4 is only about $1/2$ to that by QIF and GMMAI2 methods, and the SD of $\hat \beta_2$ by GMMAI4 method is nearly $1/3$ to that by the QIF and the relative efficiency of $\hat \beta_2$ by GMMAI4 is about $1.5$ to that by GMMAI2. Once again, these results show that applying auxiliary information effectively can help us improve the efficiency of parameter estimators.

In above simulations, we just use the first component of $\bX_1$ in making groups. In fact, it is usually very difficult to obtain the information related to all the components of covariate $\bX_1$. Thus, we conduct another simulation study to explore the relationship between $\rho_X$ and the extend of improvement in estimation efficiency when the auxiliary information is only related to part components of $\bX_1$. We estimate the parameter in model (\ref{mod}) by QIF, GMMAI2 and GMMAI4 methods, respectively when $n=300$, $\bSigma_Y$ has CS structure with $\rho_Y=0.5$, and $\bSigma_X^1$ has CS and AR(1) structures with $\rho_X=0.2,0.5$ and $0.8$. Besides of the Bias, SDs, SEs and CPs, we also calculate the relative efficiency (RE) of the estimated coefficients, which represents the variance ratio of the QIF estimator and GMM estimator with subgroup information. The results are summarized in Table \ref{Tab-3}. The table shows that the RE of $\hat \beta_1$ by the GMMAI4 method is becoming larger with the increasing of $\rho_X$. In fact, the larger $\rho_X$ is, the higher correlation among the components of $\bX_1$ is. In this case, even though the auxiliary information only be connected to $X_{11}$, it also involves the information of the other two components in $\bX_1$. Thus, the power of estimator be improved to a larger extend.

Finally, we study the impact of the auxiliary information on the power of hypothesis about parameters $\beta_1$ and $\beta_2$ in model (\ref{mod}). $\bSigma_X^1$ and $\bSigma_Y$ have CS structure with $\rho_X=0.5$ and $\rho_Y=0.5$. When $n=300$, we generate data from $\beta_1=0.5$ and $\beta_2=-0.5$. We assume that $\mathrm{H}_{01}:\beta_1=\beta_1^0$ and $\mathrm{H}_{02}:\beta_2=\beta_2^0$ with $\beta_1^0=0.5,0.55,0.6$ and $\beta_2^0=-0.5,-0.55,-0.6$. We calculate the type I errors and test power by QIF, GMMAI2 and GMMAI4 methods, respectively. Table \ref{Tab-4} is presented the simulation results. The table shows that all type I errors are close to the nominal value $0.05$, which indicates that all testing methods perform well. When incorporating subgroup information $\bphi_1^*$ and $\bphi_2^*$, the power of hypothesis $\mathrm{H}_{02}:\beta_2=\beta_2^0$ will be improved largely. The power of GMMAI2 is more than $3$ times to that of the QIF when $\beta_2^0=-0.55$. However, if we use all auxiliary information $\bphi_1-\bphi_4$ during the procedure of hypothesis test, the power of $\mathrm{H}_{01}$ and $\mathrm{H}_{02}$ will be improved at the same time. For example, when $\beta_1^0=0.55$ and $\beta_2^0=-0.60$, the power of GMMAI4 are respectively about $2$ and $3$ times to the QIF one. The results show that the auxiliary information can help us improve not only the efficiency of parameter estimation, but also the power of hypothesis test.

\begin{figure}[!htbp]
\centering
\includegraphics[width=.30\textwidth]{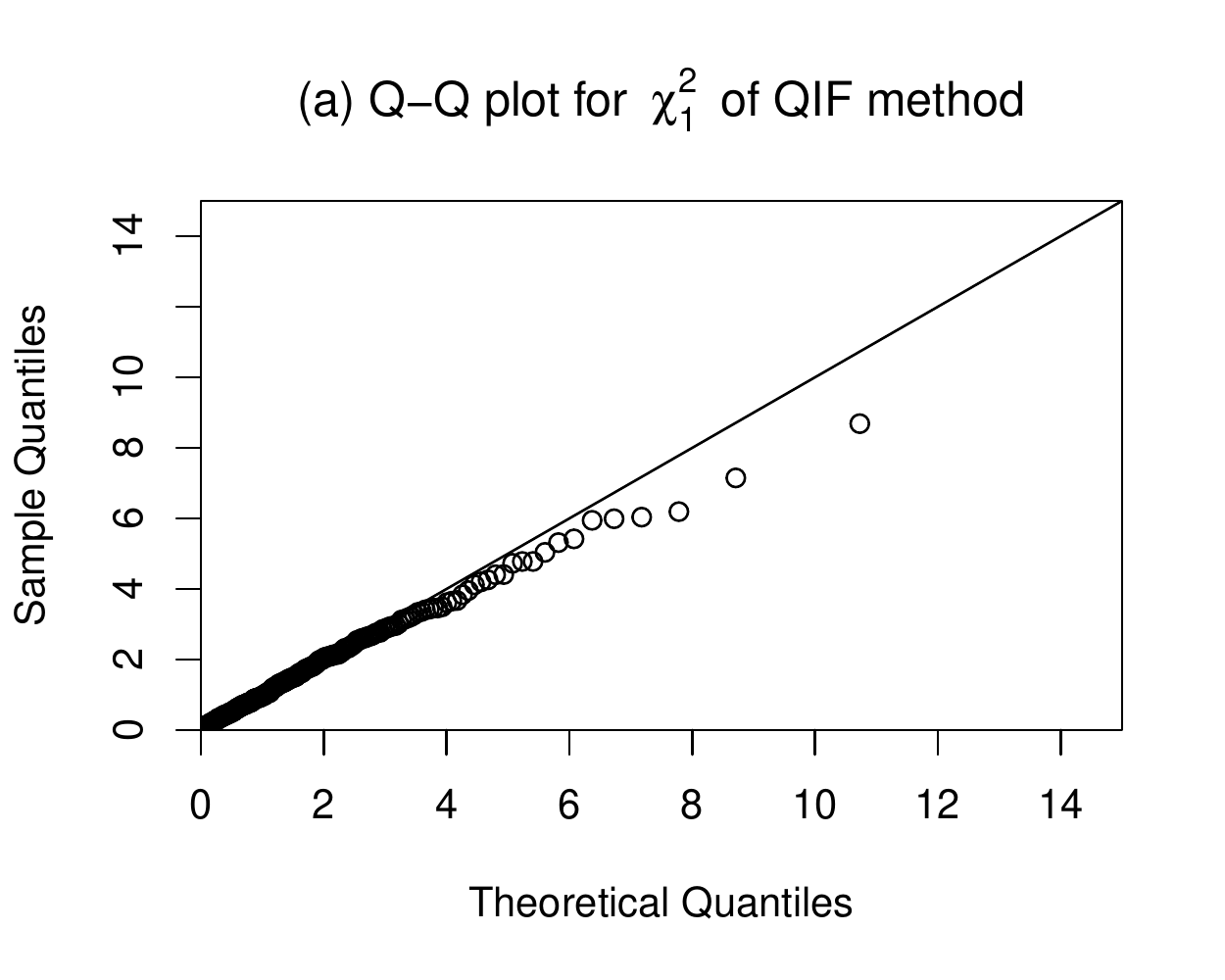}
\includegraphics[width=.30\textwidth]{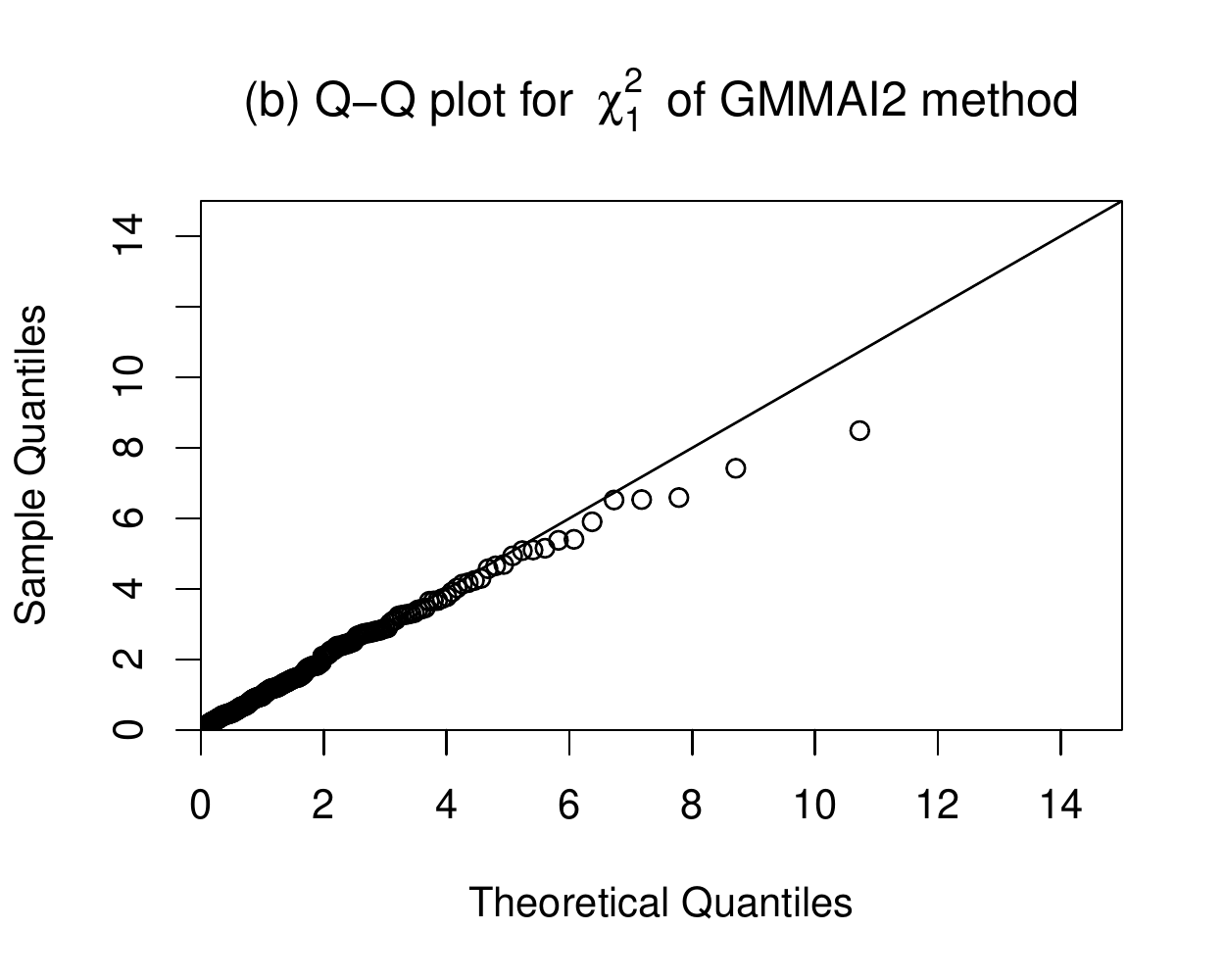}
\includegraphics[width=.30\textwidth]{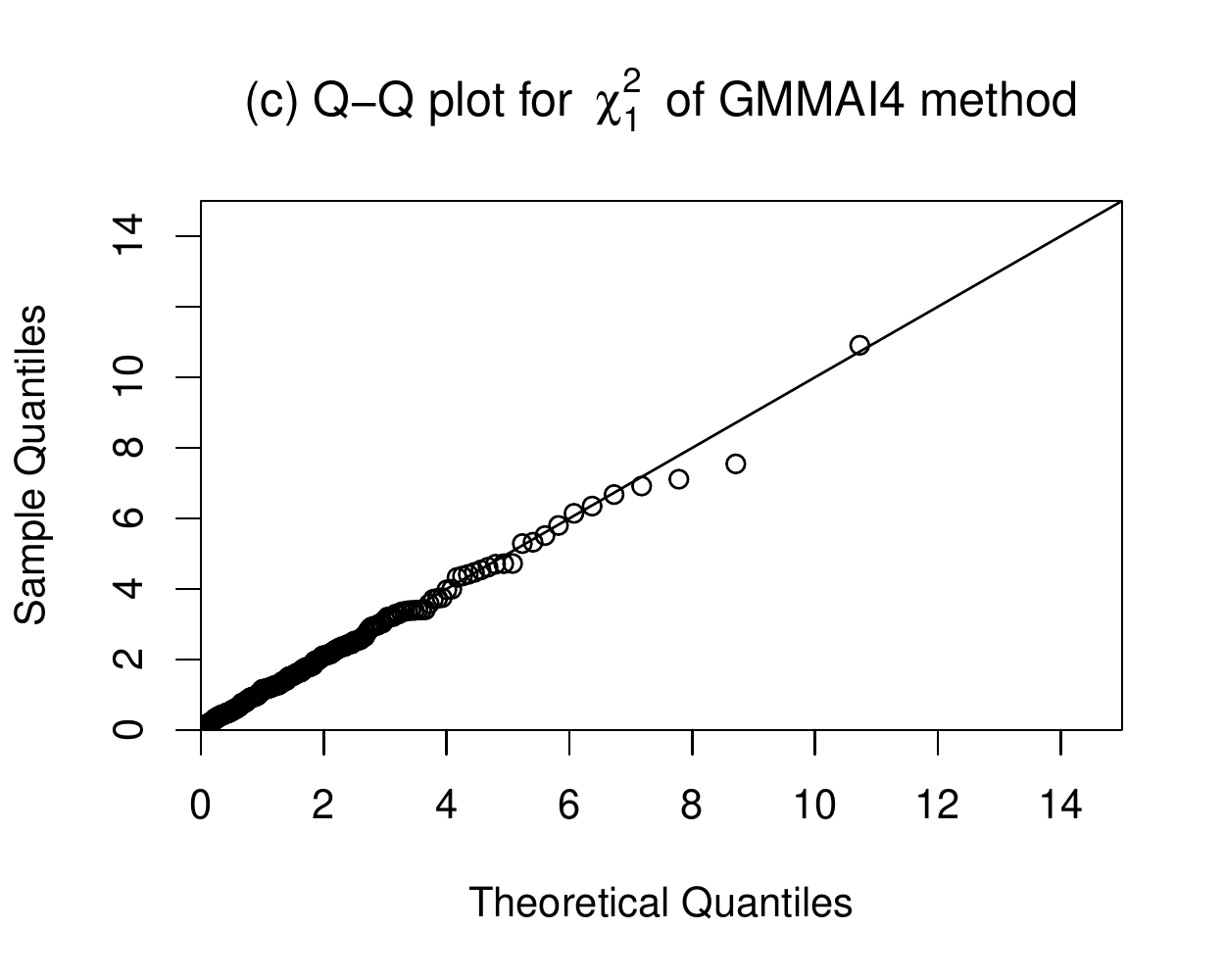}
\caption{The QQ plot of hypothesis test $\mathrm{H_{01}}: \beta_1=0.50$ by three methods.}\label{QQ-1}
\end{figure}

\begin{figure}[!htbp]
\centering
\includegraphics[width=.30\textwidth]{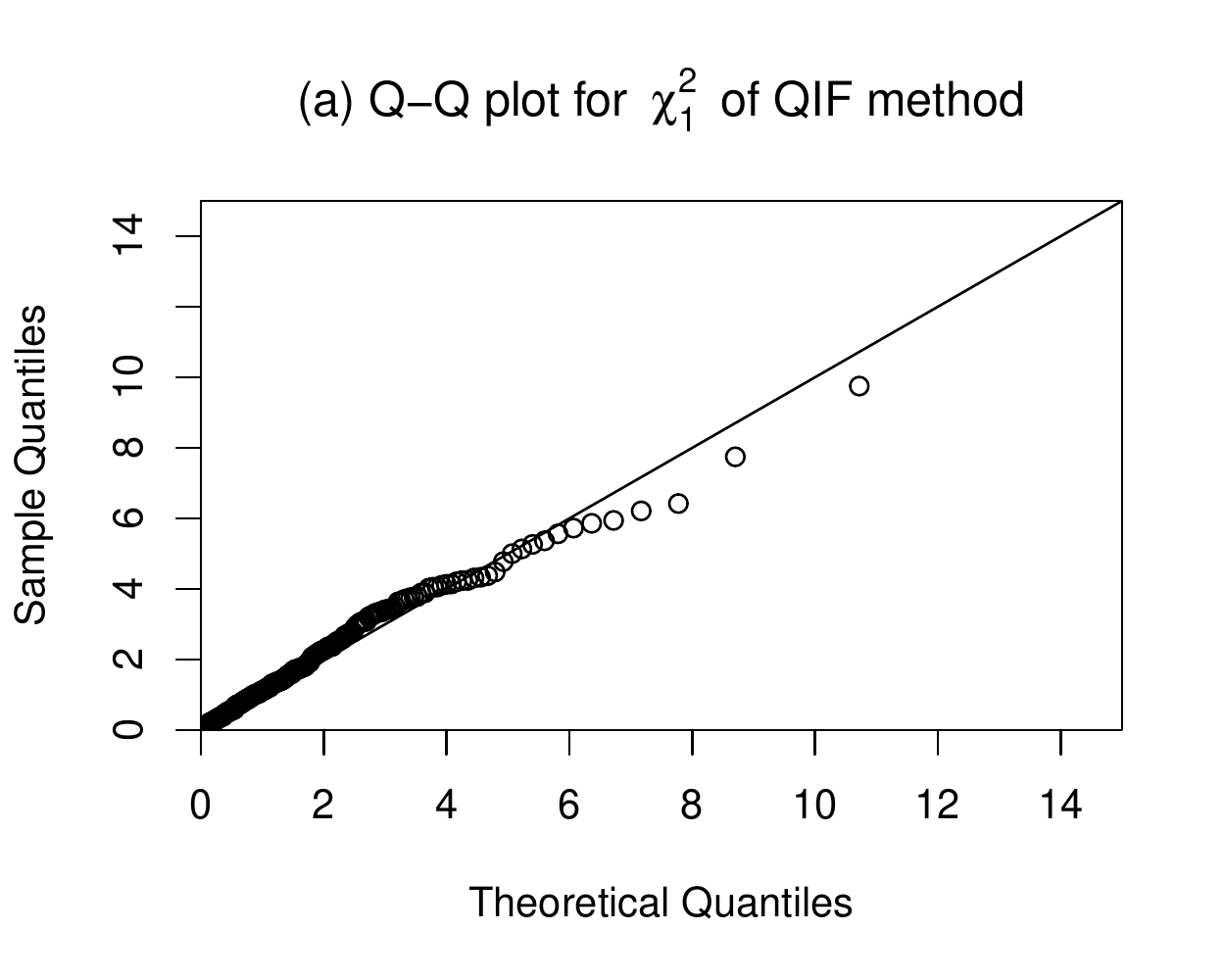}
\includegraphics[width=.30\textwidth]{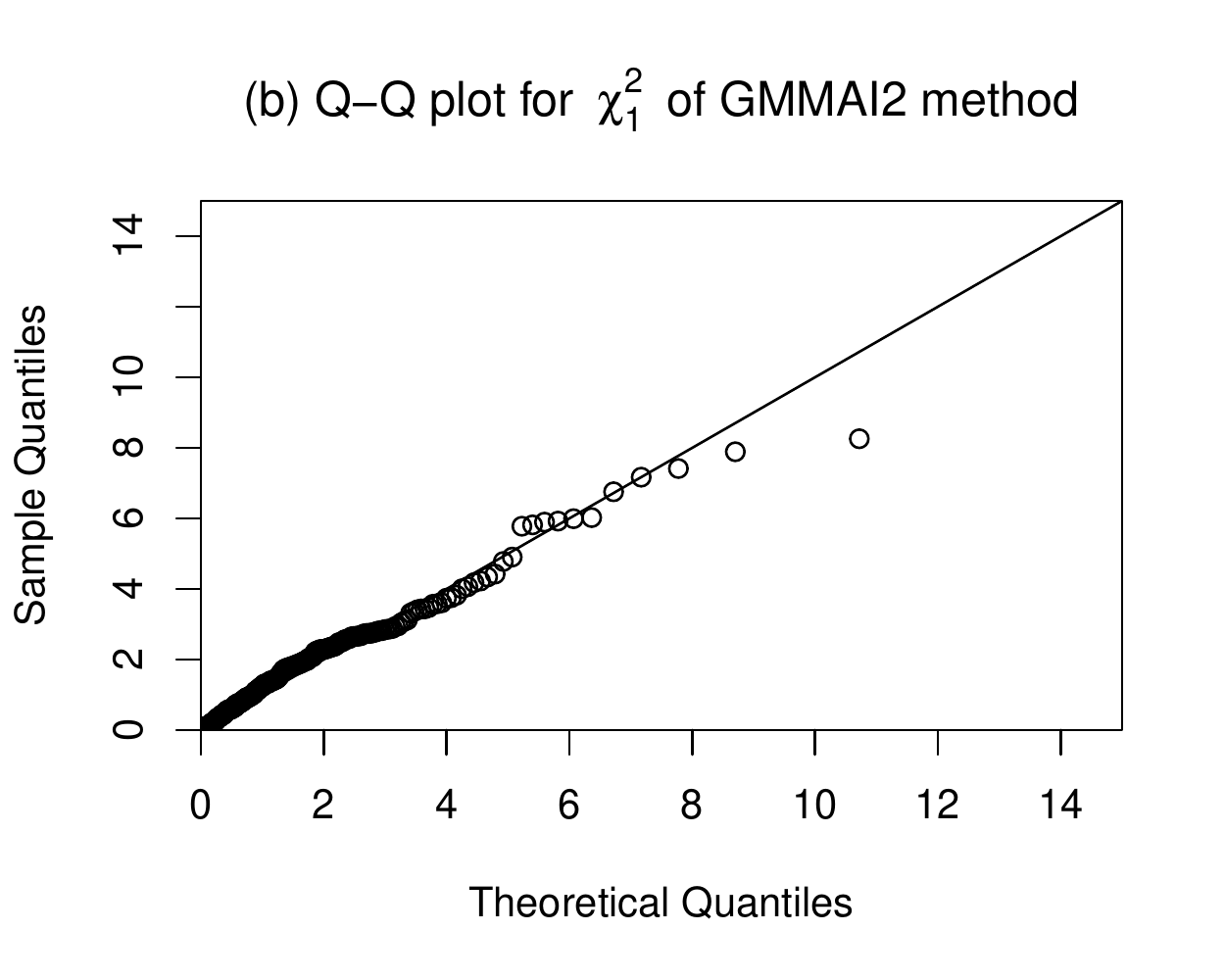}
\includegraphics[width=.30\textwidth]{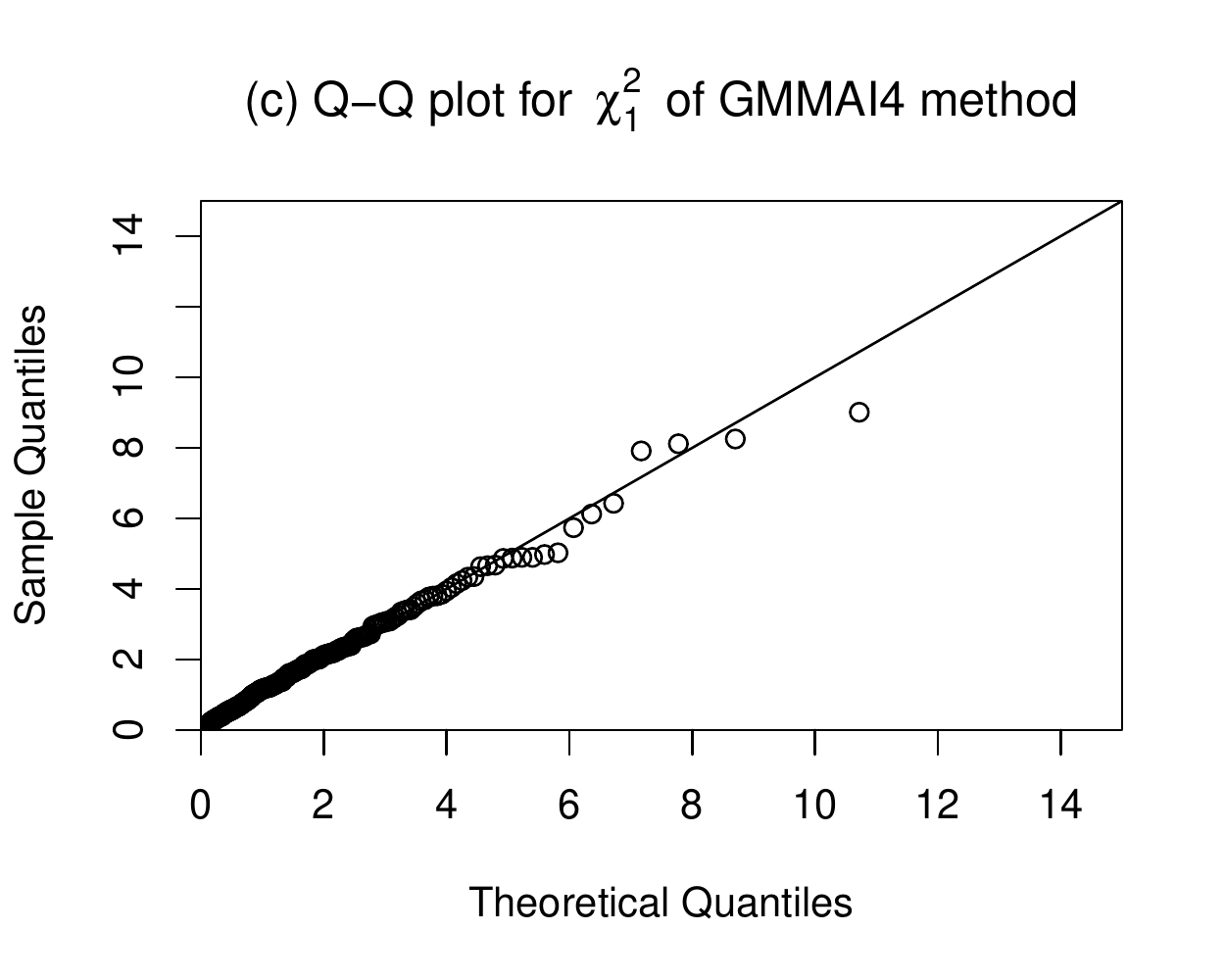}
\caption{The QQ plot of hypothesis test $\mathrm{H_{02}}: \beta_2=-0.50$ by three methods.}\label{QQ-2}
\end{figure}

In order to examine the conclusion in Theorem \ref{thm2}, we plot the QQ-plot of $n\{Q\left(\beta_1^0,\tilde{\beta}_2\right)$ $-Q\left(\hat{\beta}_1,\hat{\beta_2}\right)\}$ and $n\left\{Q\left(\tilde{\beta}_1,\beta_2^0\right)-Q\left(\hat{\beta}_1,\hat{\beta_2}\right)\right\}$ under $\beta_1^0=0.5$ and $\beta_2=-0.5$ by QIF, GMMAI2 and GMMAI4 methods, respectively, which are presented in Figure \ref{QQ-1} and \ref{QQ-2}. In these figures, the sample quantiles show linear relationship with the theoretical ones, which is consistent with the theoretical conclusion in Theorem \ref{thm2}.

\section{Real Data Analysis}

As an illustration, we applied the proposed methods to the Early Childhood Longitudinal Study, Kindergarten Class of 1998-99 (ECLS-K) database, which contains the gender and longitudinal observations of $21409$ children's reading, math and science ability scores at seven time points. Those ability scores were obtained from the Item Response Theory (IRT) study, which can be used to measure a child's underlying ability. We study the influence of reading ability, math ability and gender on children's science ability through the observations measured at Grade $3,5$ and $8$. Deleting the subjects with missing data, the sample size is $8591$. For the convenience of analysis, we standardize all the ability scores firstly. Letting the science ability score be the response, and gender, reading as well as math ability scores be covariates, which are denoted by $Y_{ij}$, $X_{i1}$(1-male,0-female), $X_{ij2}$ and $X_{ij3}$ respectively for $i=1,\cdots,8591$; $j=1,2,3$. By correlation analysis, we found that different components of $\bY$ are highly correlated. So we consider the following marginal model,
\begin{equation*}
\mathrm{E}\left(Y_{ij}\mid \bX_{ij}=\bx_{ij}\right)=\beta_1x_{i1}+\beta_2x_{ij2}+\beta_3x_{ij3}.
\end{equation*}
To obtain the auxiliary information, we divide the subjects into subgroups. Here we try three kinds of grouping manners. First of all, we group the data by the value of gender and math ability scores in Grade $3$. The subgroups can be written as
\begin{eqnarray*}
&&\Omega_1^{\mathrm{I}}=\left\{\left(X_1,\bX_3\right):X_1=1,X_{31}\geq0\right\},\quad\Omega_2^{\mathrm{I}}=\left\{\left(X_1,\bX_3\right):X_1=1,X_{31}<0\right\},\\
&&\Omega_3^{\mathrm{I}}=\left\{\left(X_1,\bX_3\right):X_1=0,X_{31}\geq0\right\},\quad\Omega_4^{\mathrm{I}}=\left\{\left(X_1,\bX_3\right):X_1=0,X_{31}<0\right\}.
\end{eqnarray*}
Combing $\Omega_1^{I}$ and $\Omega_2^{I}$, $\Omega_3^{I}$ and $\Omega_4^{I}$ respectively, we obtain two subgroups that are only related to the value of $X_1$, denoted as $\Omega_1^{\mathrm{I}*}$ and  $\Omega_2^{\mathrm{I}*}$. In the second case, we separating the subjects into groups by the reading and math ability scores in Grade $3$. The corresponding subgroups are
\begin{eqnarray*}
&&\Omega_1^{\mathrm{II}}=\left\{\left(\bX_2,\bX_3\right):X_{21}\geq0,X_{31}\geq0\right\},\quad\Omega_2^{\mathrm{II}}=\left\{\left(\bX_2,\bX_3\right):X_{21}<0,X_{31}\geq0\right\},\\
&&\Omega_3^{\mathrm{II}}=\left\{\left(\bX_2,\bX_3\right):X_{21}\geq0,X_{31}<0\right\},\quad\Omega_4^{\mathrm{II}}=\left\{\left(\bX_2,\bX_3\right):X_{21}<0,X_{31}<0\right\}.
\end{eqnarray*}
Similarly, we could obtain two subgroups $\Omega_1^{\mathrm{II}*}$ and $\Omega_2^{\mathrm{II}*}$, which are only connected with $X_{31}$, by merging $\Omega_1^{\mathrm{II}}$ and $\Omega_2^{\mathrm{II}}$, as well as $\Omega_3^{\mathrm{II}}$ and $\Omega_4^{\mathrm{II}}$ respectively. Finally, we try to use the reading ability scores in Grade $3$ and $8$ to make groups. The obtained subgroups take the forms of
\begin{eqnarray*}
&&\Omega_1^{\mathrm{III}}=\left\{\left(\bX_2\right):X_{21}\geq0,X_{23}\geq0\right\},\quad\Omega_2^{\mathrm{III}}=\left\{\left(\bX_2\right):X_{21}\geq0,X_{23}<0\right\},\\
&&\Omega_3^{\mathrm{III}}=\left\{\left(\bX_2\right):X_{21}<0,X_{23}\geq0\right\},\quad\Omega_4^{\mathrm{III}}=\left\{\left(\bX_2\right):X_{21}<0,X_{23}<0\right\}.
\end{eqnarray*}
If we incorporate $\Omega_1^{\mathrm{III}}$ and $\Omega_2^{\mathrm{III}}$, $\Omega_3^{\mathrm{III}}$ and $\Omega_4^{\mathrm{III}}$ respectively, two groups based on the value of $X_{21}$ are obtained, which can be written as $\Omega_1^{\mathrm{III}*}$ and $\Omega_2^{\mathrm{III}*}$. In order to use the proposed method, we randomly sample a subset of sample size $1000$ from the original complete data for our analysis, and the left data are used for estimating the auxiliary information. We estimate the parameters in marginal model by QIF, GMMAI2 and GMMAI4 methods under different grouping manners. In each case, ``GMMAI2" represents the GMM estimator with $2$ subgroups information and ``GMMAI4" stands for GMM estimator with $4$ subgroups information. For example, ``GMMAI2(I)" stands for GMM estimator with auxiliary information provided by subgroups $\Omega_1^{*\mathrm{I}}-\Omega_2^{*\mathrm{I}}$. The analysis results are presented in Table \ref{Tab-5}.

The table shows that the estimates of parameters by different methods are very similar. $\hat \beta_1$ are smaller than $0$, which indicates that, to children with similar reading and math ability, a girl's science ability is higher than a boy's. As we all known, the development of girl's intelligence is earlier than that of boy's, so this result is reasonable. Moreover, we observe that the estimated coefficients of reading and math ability are larger than $0$, which illustrates that these two kinds of ability have positive effects on children's science ability. A good reading ability could help children understand new things easily, while outstanding math ability does good to cultivate children's logical thinking ability. So all of them are helpful to promote children's science ability. Finally, the SEs of the estimated parameters obtained by the GMM method with subgroup information are smaller than that by the QIF, which illustrates that the auxiliary information can improve the estimation efficiency. This result is consistent with our theoretical results.

\section{Discussion}

In this paper, in order to improve the efficiency of the estimated coefficient in the marginal model, we proposed a kind of GMM procedure with auxiliary information. The asymptotic properties of the proposed estimators have been established. The simulation studies and real data analysis show that our proposed estimators are more efficient than the one obtained by the conventional QIF method. However, we just considered the application of the auxiliary information in marginal model based on a complete data set. It is of interest to explore how to apply the auxiliary information to some more complicated cases, such as the missing data, in further. In addition, we just consider the case that the auxiliary information is consistent with the sample data set we researched. It is meaningful to study how to use the auxiliary information, which is inconsistent with the data set we are interested in.

\vspace{1in}
\noindent {\Large \bf Appendix}\\
For a vector or matrix $\mathbf{v}$, $\|\mathbf{v}\|$ denotes the $L_2$-norm of $\mathbf{v}$. We impose the following regularity conditions that are needed to establish the asymptotic properties of the estimators. Throughout,  ``$\stackrel{\cal{P}}{\longrightarrow}$'' represents converge in probability.

\begin{enumerate}
\item[C1.] There exists a unique $\bbeta_0$ in a compact space, which satisfies ${\rm E}\left\{\bg(\bbeta_0,\bX)\right\}=\mathbf{0}$.

\item[C2.] ${\rm E}\left\{\bS(\bbeta_0,\bX)\bS(\bbeta_0,\bX)^T\right\}$ and ${\rm E}\left\{\bPsi(\bbeta_0,\bX)\bPsi(\bbeta_0,\bX)^T\right\}$ are positive definite and finite.

\item[C3.] Matrix-valued function $\bA(\bbeta)$ is second continuously differentiable with respect to $\bbeta$ and is uniformly bounded up to the second order partial derivatives, where $\bA(\bbeta)$ is a diagonal matrix with each entry as the marginal conditional variance of the response, ${\rm Var} (Y_j|\bX_{j}=\bx_j),j=1,\cdots,q$.

\item[C4.] The matrix $\bSigma_n(\bbeta)$ is second continuously differentiable with respect to $\bbeta$, and there exist a matrix $\bSigma(\bbeta)$ which is continuous and positive definite at $\bbeta_0$ such that
    \begin{equation*}
    \sup_{\bbeta\in V_{\varepsilon,\bbeta_0}}\|\bSigma_n(\bbeta)-\bSigma(\bbeta)\|=o_p(1),
    \end{equation*}
    where $V_{\varepsilon,\bbeta_0}$ is a neighborhood of $\bbeta_0$.
\item[C5.] The vector valued function $\bg_{(n)}(\bbeta)$ is second continuously differentiable with respect to $\bbeta$, and there exist a matrix $\bg(\bbeta)$ which is continuous at $\bbeta_0$ such that
    \begin{equation*}
    \sup_{\bbeta\in V_{\varepsilon,\bbeta_0}}\|\bg_{(n)}(\bbeta)-\bg(\bbeta)\|=o_p(1).
    \end{equation*}
\end{enumerate}
We define
\begin{eqnarray}
&\bG_n(\bbeta)&=\dfrac{\partial \bg_{(n)}(\bbeta)}{\partial \bbeta^T}=\begin{pmatrix} \dfrac{\partial \bS_{(n)}(\bbeta)}{\partial \bbeta^T} \\ \dfrac{\partial \bPsi_{(n)}(\bbeta)}{\partial \bbeta^T} \end{pmatrix},\label{GN} \\
&\bW^{(l)}(\bmu_i,\bbeta)&=A_i^{-\frac{1}{2}}\bM_l A_i^{-\frac{1}{2}}=\begin{pmatrix} w_{i11}^{(l)}(\bmu_i,\bbeta) & \cdots & w_{i1q}^{(l)}(\bmu_i,\bbeta) \\ \vdots  &   &  \vdots  \\ w_{iq1}^{(l)}(\bmu_i,\bbeta) & \cdots & w_{iqq}^{(l)}(\bmu_i,\bbeta) \end{pmatrix}, l=1,\cdots,L, \nonumber \\
&\bPhi_{n}(\bbeta)&={\rm diag} \left(\bg_{(n)}(\bbeta)^T,\cdots,\bg_{(n)}(\bbeta)^T\right),\nonumber\\
&\widetilde{\bOmega}_n(\bbeta)&={\rm diag}\left(\dfrac{\partial \bSigma_n^{-1}(\bbeta)}{\partial \beta_1},\cdots,\dfrac{\partial \bSigma_n^{-1}(\bbeta)}{\partial \beta_p}\right),\nonumber \\
{\rm and}\nonumber \\
&\bGamma_n(\bbeta)&=\left(\bg_{(n)}(\bbeta)^T,\cdots,\bg_{(n)}(\bbeta)^T\right)^T\label{Gamma}.
\end{eqnarray}

\noindent {\em Proof of Theorem \ref{thm1}}.
First, we establish the asymptotic property of the derivative of $\bS_{(n)}(\bbeta)$ with respect to $\bbeta^T$, when $\bbeta=\bbeta_0$. From the definition of $\bS_n^{(l)}(\bbeta)$, we have
\begin{equation*}
\bS_n^{(l)}(\bbeta)=\begin{pmatrix} \dfrac{1}{n}\sum\limits_{i=1}^n \sum\limits_{j=1}^q \sum\limits_{k=1}^q \dfrac{\partial \mu_{ij}}{\partial \beta_1} w_{ijk}^{(l)}(\bmu_i,\bbeta)(Y_{ik}-\mu_{ik})\\ \vdots \\ \dfrac{1}{n}\sum\limits_{i=1}^n \sum\limits_{j=1}^q \sum\limits_{k=1}^q \dfrac{\partial \mu_{ij}}{\partial \beta_p} w_{ijk}^{(l)}(\bmu_i,\bbeta)(Y_{ik}-\mu_{ik}) \end{pmatrix}.
\end{equation*}
The partial derivative of the $m$th component of $\bS_n^{(l)}(\bbeta)$ with respect to the $h$th component of $\bbeta$, $\beta_h$ can be decomposed as
\begin{equation*}
\dfrac{\partial \bS_{n_m}^{(l)}(\bbeta)}{\partial \beta_h}=a_n(\bbeta)+b_n(\bbeta)-c_n(\bbeta),
\end{equation*}
where
\begin{eqnarray*}
a_n(\bbeta)&=&\dfrac{1}{n}\sum\limits_{i=1}^n \sum\limits_{j=1}^q \sum\limits_{k=1}^q \dfrac{\partial^2 \mu_{ij}}{\partial \beta_m \partial \beta_h} w_{ijk}^{(l)}(\bmu_i, \bbeta)(Y_{ik}-\mu_{ik}), \\
b_n(\bbeta)&=&\dfrac{1}{n}\sum\limits_{i=1}^n \sum\limits_{j=1}^q \sum\limits_{k=1}^q \dfrac{\partial \mu_{ij}}{\partial \beta_m}\dfrac{\partial w_{ijk}^{(l)}(\bmu_i,\bbeta)}{\partial \beta_h}(Y_{ik}-\mu_{ik}),\\
c_n(\bbeta)&=&\dfrac{1}{n}\sum\limits_{i=1}^n \sum\limits_{j=1}^q \sum\limits_{k=1}^q \dfrac{\partial \mu_{ij}}{\partial \beta_m} w_{ijk}^{(l)}(\bmu_i, \bbeta) \dfrac{\partial \mu_{ik}}{\partial \beta_h}.
\end{eqnarray*}
From the law of large numbers and double expectation, it follows that $a_n(\bbeta_0)=o_p(1)$ and $b_n(\bbeta_0)=o_p(1)$ as $n \rightarrow \infty$.
By Slutsky's theorem, we have
\begin{equation}\label{B1}
{\dfrac{\partial \bS_{(n)}(\bbeta)}{\partial \bbeta^T}}\Big|_{\bbeta=\bbeta_0} \stackrel{\cal{P}}{\longrightarrow} \bB_1,
\end{equation}
where $\bB_1$ is a $pL \times p$ matrix, whose the $\left\{(l-1)p+m\right\}$th row and $h$th column takes the form of
\begin{equation*}
E \left\{-\sum\limits_{j=1}^q \sum\limits_{k=1}^q \dfrac{\partial \mu_{ij}}{\partial \beta_m} w_{ijk}^{(l)}(\bmu_i, \bbeta) \dfrac{\partial \mu_{ik}}{\partial \beta_h}\right\}.
\end{equation*}
Then, we consider the asymptotic property of the derivative of $\partial \bPsi_{(n)}(\bbeta) / \partial \bbeta^T$, when $\bbeta=\bbeta_0$. Note that
\begin{equation*}
\dfrac{\partial \bPsi_{(n)}(\bbeta)}{\partial \bbeta^T}=\begin{pmatrix}
\dfrac{\partial \bPsi_n^{(1)}(\bbeta)}{\partial \beta_1} & \cdots & \dfrac{\partial \bPsi_n^{(1)}(\bbeta)}{\partial \beta_p}\\ \vdots & & \vdots \\
\dfrac{\partial \bPsi_n^{(K)}(\bbeta)}{\partial \beta_1} & \cdots & \dfrac{\partial \bPsi_n^{(K)}(\bbeta)}{\partial \beta_p} \end{pmatrix},
\end{equation*}
where
\begin{equation*}
\dfrac{\partial \bPsi_n^{(k)}(\bbeta)}{\partial \beta_h}=\dfrac{1}{n}\sum\limits_{i=1}^n I(\bX_i \in \Omega_k)\dfrac{\partial \bmu_i}{\partial \beta_h}.
\end{equation*}
By Slutsky's theorem, we obtain
\begin{equation*}
\dfrac{\partial \bPsi_{(n)}(\bbeta)}{\partial \bbeta^T}\Big|_{\bbeta=\bbeta_0}\stackrel{\cal{P}}{\longrightarrow}\bB_2,
\end{equation*}
where
\begin{equation}\label{B2}
\bB_2={\begin{pmatrix} \bb_{11} & \cdots & \bb_{1p} \\ \vdots & & \vdots \\ \bb_{K1} & \cdots & \bb_{Kp}\end{pmatrix}}_{Kq \times p}
\end{equation} and $\bb_{mh}=E \left\{I(\bX_i \in \Omega_m) \dfrac{\partial \bmu_i}{\partial \beta_h} |_{\bbeta=\bbeta_0}\right\}, m=1,\cdots,K; h=1,\cdots,p$. Combining (\ref{GN}), (\ref{B1}) and (\ref{B2}), one can show that
\begin{equation*}
\bG_n(\bbeta_0)\stackrel{\cal{P}}{\longrightarrow}\bB,
\end{equation*}
where $\bB=\begin{pmatrix} \bB_1^T,&\bB_2^T \end{pmatrix}^T$.

Noted that $Q_n(\bbeta)=\bg_{(n)}(\bbeta)^T\bSigma_n^{-1}(\bbeta)\bg_{(n)}(\bbeta)$, the estimating equation takes the form of
\begin{equation*}
\bU_n(\bbeta)=2\bG_n(\bbeta)^T\bSigma_n^{-1}(\bbeta)\bg_{(n)}(\bbeta)+\bPhi_n(\bbeta) \widetilde{\bOmega}_n(\bbeta)\bGamma_n(\bbeta).
\end{equation*}
From Slutsky's theorem, we have that $\bU_n(\bbeta_0)\stackrel{\cal{P}}{\longrightarrow}{\mathbf 0}$. Thus, $\widehat{\bbeta}_{\rm GMM}  \stackrel{\cal{P}}{\longrightarrow} \bbeta_0$.

By Taylor expansion, we have
\begin{equation*}
\widehat{\bbeta}_{\rm GMM}-\bbeta_0=\left\{-\frac{\partial \bU_n(\bbeta)}{\partial \bbeta}\Big|_{\bbeta=\bbeta_0}\right\}^{-1}\bU_n(\bbeta_0)+o_p(1).
\end{equation*}
Noting
\begin{equation*}
\frac{\partial \bU_n(\bbeta)}{\partial \bbeta}\Big|_{\bbeta=\bbeta_0}=2\bG_n(\bbeta_0)^T\bSigma_n^{-1}(\bbeta_0)\bG_n(\bbeta_0)+o_p(1),
\end{equation*}
we obtain
\begin{eqnarray}\label{expression}
n^{1/2}(\widehat{\bbeta}_{\rm GMM}-\bbeta_0)&=&\{
-\frac{\partial \bU_n(\bbeta)}{\partial \bbeta}\Big|_{\bbeta=\bbeta_0}\}^{-1}\{n^{1/2}\bU_n(\bbeta_0)\}+o_p(1)
\notag\\
&=&-[\{\bG_n(\bbeta_0)^T\bSigma_n^{-1}(\bbeta_0)\bG_n(\bbeta_0)\}^{-1}\bG_n(\bbeta_0)^ T\bSigma_n^{-1}(\bbeta_0)]\{n^{1/2}\bg_{(n)}(\bbeta_0)\}+o_p(1).
\notag\\
\end{eqnarray}
By the law of large numbers, we obtain
\begin{equation*}
\bSigma_n(\bbeta_0)\stackrel{\cal{P}}{\longrightarrow}E\left\{\bg(\bbeta_0,\bX_i)\bg(\bbeta_0,\bX_i)^ T\right\}
=\begin{pmatrix} E\left\{\bS(\bbeta_0,\bX_i)\bS(\bbeta_0,\bX_i)^T\right\} & E\left\{\bS(\bbeta_0,\bX_i)\bPsi(\bbeta_0,\bX_i)^T\right\} \\
E\left\{\bPsi(\bbeta_0,\bX_i)\bS(\bbeta_0,\bX_i)^T\right\} & E\left\{\bPsi(\bbeta_0,\bX_i)\bPsi(\bbeta_0,\bX_i)^T\right\} \end{pmatrix}.
\end{equation*}
Let $\bSigma_1=E\left\{\bS(\bbeta_0,\bX_i)\bS(\bbeta_0,\bX_i)^T\right\}$ and $\bSigma_2=E\left\{\bPsi(\bbeta_0,\bX_i)\bPsi(\bbeta_0,\bX_i)^T\right\}$. Using double expectation, we obtain
\begin{eqnarray*}
E\left\{\bS(\bbeta_0,\bX_i)\bPsi(\bbeta_0,\bX_i)^T\right\}&=&E\left[E\left\{\bS(\bbeta_0,\bX_i)\bPsi(\bbeta_0,\bX_i)^T|\bX_i\right\}\right]\\
&=&E[E\left\{\bS(\bbeta_0,\bX_i)|\bX_i\right\}\bPsi(\bbeta_0,\bX_i)^T]=\mathbf{0}.
\end{eqnarray*}
Similarly, we have $E\left\{\bPsi(\bbeta_0,\bX_i)\bS(\bbeta_0,\bX_i)^T\right\}=\mathbf{0}$. Thus,
\begin{equation*}
\bSigma_n(\bbeta_0)\stackrel{\cal{P}}{\longrightarrow}\bSigma_0,
\end{equation*}
where $\bSigma_0={\rm diag}(\bSigma_1, \bSigma_2 )$.

By central limit theorem, we observe
\begin{equation*}
n^{1/2}\bg_{(n)}({\bbeta_0})\stackrel{\cal{D}}{\longrightarrow}N_{pL+Kq}(\mathbf{0},\bSigma_0).
\end{equation*}
From (\ref{expression}) and the sandwich formula, we have
\begin{equation*}
n^{1/2}(\widehat{\bbeta}_{\rm GMM}-\bbeta_0)\stackrel{\cal{D}}{\longrightarrow} N(\mathbf{0},\widetilde{\bSigma}_0),
\end{equation*}
where
$
\widetilde{\bSigma}_0=\left(\bB^T\bSigma_0^{-1}\bB\right)^{-1}=\left(\bB_1^T\bSigma_1^{-1}\bB_1+\bB_2^T\bSigma_2^{-1}\bB_2\right)^{-1}.
$
This completes the proof of Theorem \ref{thm1}.
\qed

\noindent {\em Proof of Theorem \ref{thm2}}.
During the proof, we delete the subscript ``GMM" from $\widehat{\bgamma}_{\rm GMM}$ for simplicity. Denote
\begin{eqnarray*}
&&\dfrac{\partial Q_n}{\partial \bgamma}=\dot{Q}_\gamma, \frac{\partial Q_n}{\partial \blambda}=\dot{Q}_\lambda, \dfrac{\partial^2 Q_n}{\partial \bgamma \partial \bgamma^T}=\ddot{Q}_{\gamma\gamma},
\dfrac{\partial^2 Q_n}{\partial \blambda \partial \blambda^T}=\ddot{Q}_{\lambda\lambda},
\dfrac{\partial^2 Q_n}{\partial \bgamma \partial \blambda^T}=\ddot{Q}_{\gamma\lambda} \quad{\rm and}\\
&&\dfrac{\partial^2 Q_n}{\partial \blambda \partial \bgamma^T}=\ddot{Q}_{\lambda\gamma}.
\end{eqnarray*}
By Taylor expansion, we have
\begin{equation}\label{eq1}
Q_n(\bgamma_0,\blambda_0)-Q_n(\widehat{\bgamma},\widehat{\blambda})=\frac{1}{2}{\begin{pmatrix} \bgamma_0-\widehat{\bgamma} \\ \blambda_0-\widehat{\blambda} \end{pmatrix}}^T
\begin{pmatrix} {\ddot{Q}}_{\gamma\gamma}(\bgamma^*,\blambda^*) & {\ddot{Q}}_{\gamma\lambda}(\bgamma^*,\blambda^*) \\ {\ddot{Q}}_{\lambda\gamma}(\bgamma^*,\blambda^*)& {\ddot{Q}}_{\lambda\lambda}(\bgamma^*,\blambda^*)\end{pmatrix}
\begin{pmatrix} \bgamma_0-\widehat{\bgamma} \\ \blambda_0-\widehat{\blambda} \end{pmatrix},
\end{equation}
where $(\bgamma^*,\blambda^*)$ is a point on the line segment connecting $(\bgamma_0,\blambda_0)$ and  $(\widehat{\bgamma},\widehat{\blambda})$, and
\begin{equation}\label{eq2}
Q_n(\bgamma_0,\blambda_0)-Q_n(\bgamma_0,\widetilde{\blambda})=\frac{1}{2}
{\begin{pmatrix} \mathbf{0} \\ \blambda_0-\widetilde{\blambda} \end{pmatrix}}^T
\begin{pmatrix} {\ddot{Q}}_{\gamma\gamma}(\bgamma_0,\blambda^\dag) & {\ddot{Q}}_{\gamma\lambda}(\bgamma_0,\blambda^\dag) \\ {\ddot{Q}}_{\lambda\gamma}(\bgamma_0,\blambda^\dag) & {\ddot{Q}}_{\lambda\lambda}(\bgamma_0,\blambda^\dag) \end{pmatrix}
\begin{pmatrix}
\mathbf{0} \\ \blambda_0-\widetilde{\blambda} \end{pmatrix},
\end{equation}
where $(\bgamma_0,\blambda^\dag)$ is a point on the line segment connecting $(\bgamma_0,\blambda_0)$ and  $(\bgamma_0,\widetilde{\blambda})$.
\\
Let
\begin{equation*}
\ddot{\bQ}(\bgamma^*,\blambda^*)=\begin{pmatrix} {\ddot{Q}}_{\gamma\gamma}(\bgamma^*,\blambda^*) & {\ddot{Q}}_{\gamma\lambda}(\bgamma^*,\blambda^*) \\ {\ddot{Q}}_{\lambda\gamma}(\bgamma^*,\blambda^*)& {\ddot{Q}}_{\lambda\lambda}(\bgamma^*,\blambda^*)\end{pmatrix}\quad{\rm and}\quad
\ddot{\bQ}(\bgamma_0,\blambda^\dag)=\begin{pmatrix} {\ddot{Q}}_{\gamma\gamma}(\bgamma_0,\blambda^\dag) & {\ddot{Q}}_{\gamma\lambda}(\bgamma_0,\blambda^\dag) \\ {\ddot{Q}}_{\lambda\gamma}(\bgamma_0,\blambda^\dag) & {\ddot{Q}}_{\lambda\lambda}(\bgamma_0,\blambda^\dag) \end{pmatrix}.
\end{equation*}
From (\ref{eq1}) and (\ref{eq2}), we have
\begin{equation*}
Q(\bgamma_0,\widetilde{\blambda})-Q(\widehat{\bgamma},\widehat{\blambda})
=\dfrac{1}{2}{\begin{pmatrix} \widehat{\bgamma}-\bgamma_0 \\\widehat{\blambda}- \blambda_0 \end{pmatrix}}^T
\ddot{\bQ}(\bgamma^*,\blambda^*) \begin{pmatrix} \bgamma_0-\widehat{\bgamma} \\ \blambda_0-\widehat{\blambda} \end{pmatrix}-
\frac{1}{2}
{\begin{pmatrix} \mathbf{0} \\\widetilde{\blambda}- \blambda_0 \end{pmatrix}}^T
\ddot{\bQ}(\bgamma_0,\blambda^\dag) \begin{pmatrix}
\mathbf{0} \\ \widetilde{\blambda}-\blambda_0 \end{pmatrix}.
\end{equation*}
By Taylor expansion, it follows that
\begin{eqnarray*}
\mathbf{0}&=&\dot{Q}_\lambda(\bgamma_0,\widetilde{\blambda})=\dot{Q}_\lambda(\bgamma_0,\blambda_0)+\ddot{Q}_{\lambda\lambda}
(\bgamma_0,\blambda_0)(\widetilde{\blambda}-\blambda_0)+o_p(n^{-1/2}),\\
\mathbf{0}&=&\dot{Q}_\lambda(\widehat{\bgamma},\widehat{\blambda})=\dot{Q}_\lambda(\bgamma_0,\blambda_0)+\ddot{Q}_{\lambda\lambda}
(\bgamma_0,\blambda_0)(\widehat{\blambda}-\blambda_0)+\ddot{Q}_{\lambda\gamma}
(\bgamma_0,\blambda_0)(\widehat{\bgamma}-\bgamma_0) +o_p(n^{-1/2}).
\end{eqnarray*}
Then,
\begin{equation*}
\tilde{\blambda}-\blambda_0=\ddot{Q}_{\lambda\lambda}^{-1}(\bgamma_0,\blambda_0)\ddot{Q}_{\lambda\gamma}
(\bgamma_0,\blambda_0)(\widehat{\bgamma}-\bgamma_0)+(\widehat{\blambda}-\blambda_0)+o_p(n^{-1/2}).
\end{equation*}
Thus,
\begin{equation*}
\begin{pmatrix} \mathbf{0} \\ \widetilde{\blambda}-\blambda_0 \end{pmatrix}=
\begin{pmatrix} \mathbf{0} & \mathbf{0}\\ \ddot{Q}_{\lambda\lambda}^{-1}(\bgamma_0,\blambda_0)\ddot{Q}_{\lambda\gamma}
(\bgamma_0,\blambda_0)& \mathbf{0} \end{pmatrix}
\begin{pmatrix} \widehat{\bgamma}-\bgamma_0 \\ \widehat{\blambda}-\blambda_0 \end{pmatrix}+o_p(n^{-\frac{1}{2}}).
\end{equation*}
From the proof of Theorem \ref{thm1}, we obtain
\begin{equation*}
\frac{1}{2} \begin{pmatrix}
{\ddot{Q}}_{\gamma\gamma}(\bgamma^*,\blambda^*) & {\ddot{Q}}_{\gamma\lambda}(\bgamma^*,\blambda^*) \\ {\ddot{Q}}_{\lambda\gamma}(\bgamma^*,\blambda^*)& {\ddot{Q}}_{\lambda\lambda}(\bgamma^*,\blambda^*)\end{pmatrix}
\stackrel{\cal{P}}{\longrightarrow}
\bB^T\bSigma_0^{-1}\bB,
\end{equation*}
and
\begin{equation*}
\frac{1}{2} \begin{pmatrix}
{\ddot{Q}}_{\gamma\gamma}(\bgamma_0,\blambda^\dagger) & {\ddot{Q}}_{\gamma\lambda}(\bgamma_0,\blambda^\dagger) \\ {\ddot{Q}}_{\lambda\gamma}(\bgamma_0,\blambda^\dagger)&{\ddot{Q}}_{\lambda\lambda}(\bgamma_0,\blambda^\dagger)\end{pmatrix}
\stackrel{\cal{P}}{\longrightarrow}
\bB^T\bSigma_0^{-1}\bB.
\end{equation*}
Dividing matrix $\bB$ into a 2-by-2 block matrix according to the dimensions of $\bgamma$ and $\blambda$, we have
\begin{equation*}
\bB=\begin{pmatrix} \bB_1 & \bB_2 \end{pmatrix} =\begin{pmatrix} \bB_{11} & \bB_{12} \\ \bB_{21} & \bB_{22} \end{pmatrix},
\end{equation*}
where $\bB_{11}$, $\bB_{12}$, $\bB_{21}$ and $\bB_{22}$ are $pL\times p_1$, $pL\times (p-p_1)$, $Kq\times p_1$ and $Kq\times(p-p_1)$ matrices respectively. Thus, $\bB^T\bSigma_0^{-1}\bB$ has the following form
\begin{eqnarray*}
\bB^T\bSigma_0^{-1}\bB&=&\begin{pmatrix} \bB_{11}^T & \bB_{21}^T \\ \bB_{12}^T & \bB_{22}^T \end{pmatrix}
\begin{pmatrix} \bSigma_1^{-1} & \mathbf{0} \\ \mathbf{0} & \bSigma_2^{-1} \end{pmatrix}
\begin{pmatrix} \bB_{11} & \bB_{12} \\ \bB_{21} & \bB_{22} \end{pmatrix}\\
&=&\begin{pmatrix} \bB_{11}^T\bSigma_1^{-1}\bB_{11}+\bB_{21}^T\bSigma_2^{-1}\bB_{21} & \bB_{11}^T\bSigma_1^{-1}\bB_{12}+\bB_{21}^T\bSigma_2^{-1}\bB_{22} \\
\bB_{12}^T\bSigma_1^{-1}\bB_{11}+\bB_{22}^T\bSigma_2^{-1}\bB_{21} &
\bB_{12}^T\bSigma_1^{-1}\bB_{12}+\bB_{22}^T\bSigma_2^{-1}\bB_{22} \end{pmatrix}\\
&\stackrel{\wedge}{=}&  \begin{pmatrix} \bJ_{\gamma\gamma} & \bJ_{\gamma\lambda} \\ \bJ_{\lambda\gamma} & \bJ_{\lambda\lambda} \end{pmatrix}.
\end{eqnarray*}
Therefore,
\begin{eqnarray*}
&&n\left\{Q_n(\bgamma_0,\widetilde{\blambda})-Q_n(\widehat{\gamma},\widehat{\lambda})\right\}\\
&&=n {\begin{pmatrix} \widehat{\bgamma}-\bgamma_0 \\ \widehat{\blambda}-\blambda_0 \end{pmatrix}}^T \left\{
\begin{pmatrix}
\bJ_{\gamma\gamma} & \bJ_{\gamma\lambda} \\ \bJ_{\lambda\gamma} & \bJ_{\lambda\lambda} \end{pmatrix}-
\begin{pmatrix}
\mathbf{0} & \bJ_{\gamma\lambda}\bJ_{\lambda\lambda}^{-1} \\ \mathbf{0} & \bI \end{pmatrix}
\begin{pmatrix}
\bJ_{\gamma\gamma} & \bJ_{\gamma\lambda} \\ \bJ_{\lambda\gamma} & \bJ_{\lambda\lambda} \end{pmatrix}
\begin{pmatrix}
\mathbf{0} & \mathbf{0} \\ \bJ_{\lambda\lambda}^{-1} \bJ_{\lambda\gamma} & \bI \end{pmatrix} \right\}
{\begin{pmatrix} \widehat{\bgamma}-\bgamma_0 \\ \widehat{\blambda}-\blambda_0 \end{pmatrix}}\\
&&\quad\quad+o_p(1)\\
&&=n(\widehat{\bgamma}-\bgamma_0)^T(\bJ_{\gamma\gamma}-\bJ_{\gamma\lambda}\bJ_{\lambda\lambda}^{-1}\bJ_{\lambda\gamma})
(\widehat{\bgamma}-\bgamma_0)+o_p(1).
\end{eqnarray*}
Noting the conclusion of Theorem \ref{thm1}, one can show that
\begin{equation*}
n^{1/2}(\widehat{\bgamma}-\bgamma_0) \stackrel{\cal{D}}{\longrightarrow} N_{p_1}\left\{\mathbf{0},(\bJ_{\gamma\gamma}-\bJ_{\gamma\lambda}\bJ_{\lambda\lambda}^{-1}\bJ_{\lambda\gamma})^{-1}\right\}.
\end{equation*}
Thus, the asymptotic distribution of $n\left\{Q_n(\bgamma_0,\widetilde{\blambda})-Q_n(\widehat{\gamma},\widehat{\lambda})\right\}$ is $\chi_{p_1}^2$.

This completes the proof of Theorem \ref{thm2}.
\qed



\newpage

\vspace{1in} \noindent
{\Large \bf References}

\begin{description}
\item
{ Chatterjee}, N., { Chen}, Y. H., {Maas}, P. and { Carroll}, R. (2016).
Constrained maximum likelihood estimation for model calibration using
summary-level information from external big data sources. {\it
Journal of American Statistical Association}, {\bf 111}, 107--117.


\item
{ Hansen}, L. P. (1982). Large sample properties of generalized method of moments estimators. {\it Econometrica}, {\bf 50}, 1029--1054.

\item
{ Huang}, C. Y., { Qin}, J. and { Tsai}, H. T. (2016). Efficient estimation of the Cox model with auxiliary subgroup
survival information. {\it Journal of American Statistical Association}, {\bf 111}, 787--799.

\item
{ Li}, H. and { Yin}, G. S. (2009). Generalized method of moments estimation for linear regression with
clustered failure time data. {\it Biometrika}, {\bf 96}, 293--306.

\item
{ Li}, H., { Duan}, X. G. and { Yin}, G. S. (2016). Generalized method of moments for additive hazards model with clustered dental survival data. {\it Scandinavian Journal of Statistics}, {\bf 43}, 1124--1139.

\item
{ Liang}, K. Y. and { Zeger}, S. L. (1986). Longitudinal data analysis using generalized linear models. {\it Biometrika}, {\bf 73}, 13--22.

\item
{ McCullagh}, P. and { Nelder}, J. A. (1989). Generalized Linear Models. {\em New York: Chapman and Hall}.

\item
{ Nelder}, J. A. and { Wedderburn}, R. W. M. (1972). Generalized linear models. {\it Journal of the Royal Statistical Society. Series A}, {\bf 135}, 370--384.



\item
{ Qin}, J., { Zhang}, H., { Li}, P. F., { Albanes}, D. and { Yu}, K. (2015).
Using covariate-specific disease prevalence information to increase the power
of case-control studies. {\it Biometrika}, {\bf 102}, 169--180.

\item
{ Qu}, A., { Lindsay}, B. G. and { Li}, B. (2000). Improving generalised estimating equations
using quadratic inference functions. {\it Biometrika}, {\bf 87}, 823--836.

\item
{ Qu}, A. and { Li}, R. Z. (2006). Quadratic inference functions for varying-coefficient models with longitudinal data. {\it Biometrics}, {\bf 62}, 379--391.

\end{description}

\newpage
\begin{table}[htbp]
\begin{center}
\caption{Simulation results of model (\ref{mod}) with different structures of $\bSigma_Y$ and $\bR$.}\label{Tab-1}
{
\setlength{\tabcolsep}{1.5mm}
\begin{tabular}{llll ccccccccc}
\hline
 & & & &  \multicolumn{4}{c}{$\beta_1$}&  &\multicolumn{4}{c}{$\beta_2$}  \\
\cline{5-8}\cline{10-13}
$\rho_Y$ &  $\bSigma_Y$  & WC & method & Bias & SD & SE & CP($\%$) & & Bias & SD & SE & CP($\%$)\\
\hline
0.2  & CS    & CS  & QIF  & $-$1   &  43  & 43  &	94.4  &  &	1  & 73  &	67 	& 93.2 \\
     &       &     & GMMAI& $-$1   &  45  & 42  &	93.4  &  &	1  & 36  &	35 	& 95.0 \\
     &       &AR(1)& QIF  & $-$1   &  44  & 43  &	93.4  &  &	0  & 72  &	67 	& 93.2 \\
     &       &     & GMMAI& $-$1   &  46  & 42  &	91.9  &  &	1  & 36  &	35 	& 95.0 \\
\\
     & AR(1) & CS  & QIF  &	 0 	   &  44  &	42 	&	93.6  &   & $-$5  &	65 	& 64  &	95.0\\
     &       &     & GMMAI&$-$1    &  45  &	41 	&	92.4  &   &	$-$2  &	34 	& 34  &	95.8\\
     &       &AR(1)& QIF  &$-$1    &  43  &	42 	&	93.8  &   &	$-$3  &	64 	& 64  &	94.8\\
     &       &     & GMMAI&$-$2    &  45  &	41 	&	92.2  &   &	$-$1  &	35 	& 34  &	95.0\\
\hline
0.5  & CS    & CS  & GMM  & 0 	   &  41  & 40  &	94.8  &   &	1     & 84  & 81  & 94.6 \\
     &       &     & GMMAI& 0 	   &  41  & 40  &	95.0  &   &	2     & 38  & 36  & 92.9 \\
     &       &AR(1)& GMM  & 0 	   &  43  & 42  &	94.4  &   &	1     & 84  & 81  & 94.6\\
     &       &     & GMMAI& 0 	   &  43  & 42  &	94.8  &   &	2     & 38  & 36  & 92.9\\
\\
     &AR(1)  & CS  & QIF  & 0 	   &  41  &	41 	&	94.2  &   &$-$6   &	83 	& 76  &	91.4\\
     &       &     & GMMAI&$-$1    &  41  &	41 	&	93.6  &   & 1 	  &	39 	& 35  &	92.8\\
     &       &AR(1)& QIF  &	0 	   &  39  &	40 	&	95.0  &   &$-$5   &	83 	& 75  &	92.2\\
     &       &     & GMMAI&	$-$1   &  39  &	40 	&	94.4  &   & 1 	  &	40 	& 35  &	92.6\\
\hline
0.8  & CS    & CS  & GMM  & $-$1   &  29  & 29  &	94.4  &   & $-$6  & 91  & 92  & 96.6\\
     &       &     & GMMAI& $-$1   &  30  & 29  &	94.4  &   & $-$2  & 39  & 37  & 93.0\\
     &       & AR(1)&GMM  & 0 	   &  34  & 34  &	94.6  &   & $-$7  & 91  & 92  & 95.8\\
     &       &      &GMMAI& $-$1   &  35  & 33  &	93.6  &   & $-$2  & 39  & 37  & 92.6\\
\\
     & AR(1) & CS  & QIF  &	$-$2   &  34  &	32 	&	93.6  &   & 3 	  & 90 	& 89  &	95.4\\
     &       &     & GMMAI&	$-$1   &  35  &	32 	&	93.0  &   & 1 	  &	39 	& 36  &	92.8\\
     &       &AR(1)& QIF  & $-$2   &  33  &	32 	&	93.6  &   & 4 	  &	89 	& 89  &	94.6\\
     &       &     & GMMAI&	$-$2   &  34  &	32 	&	93.2  &   & 1 	  &	39 	& 36  &	93.2\\
\hline
\end{tabular}
}
\end{center}
{\footnotesize Note: $\bSigma_Y$ represents the covariance matrix of $\bY$, WC is the structure working correlation matrix; Bias is the mean bias, SD is the standard deviation, SE is the standard error, and CP is the coverage probability, all are based on $500$ replications; QIF represents Qu et al. (2000) estimator, GMMAI represents our GMM estimator with subgroup information $\bphi_1^*$ and $\bphi_2^*$.}
\end{table}

\begin{table}[htbp]
\begin{center}
\caption{Simulation results of model (\ref{mod}) under different $\rho_Y$ with $\bSigma_X^1=\bI_3$ and sample size $n=200,500$.}\label{Tab-2}
{
\setlength{\tabcolsep}{1.0mm}
\begin{tabular}{llll ccccccccc}
\hline
  & & & & \multicolumn{4}{c}{$\beta_1$}&  &\multicolumn{4}{c}{$\beta_2$}  \\
\cline{5-8}\cline{10-13}
$\rho_Y$ & n & $\bSigma_Y$ & method & Bias & SD & SE & CP($\%$) &  & Bias & SD & SE & CP($\%$)\\
\hline
    0.2  & 200 & CS  &  QIF	    &	1 	&	40 	&	39 	&	93.8 	&  & 3 	  &	70 	&	67 	& 93.4\\
         &     &     & GMMAI2	&	1 	&	41 	&	38 	&	92.6 	&  & $-$1 &	28 	&	26 	& 93.0\\
         &     &     & GMMAI4	&	4 	&	24 	&	21 	&	94.0 	&  & 0 	  &	23 	&	21 	& 91.4\\
         &     &AR(1)& QIF	    & $-$1 	&	41 	&	39 	&	94.0 	&  & 4 	  &	67 	&	65 	& 95.2 \\
         &     &     & GMMAI2	& $-$2 	&	43 	&	39 	&	92.8 	&  & 1 	  &	28 	&	26 	& 93.2\\
         &     &     & GMMAI4	&   5 	&	24 	&	22 	&	92.6 	&  & 0 	  &	23 	&	21 	& 92.2 \\
\\
         & 500 & CS  &  QIF	    &	0 	&	25 	&	25 	&	94.2 	&  & $-$2 &	45 	&	43 	& 94.6\\
         &     &     & GMMAI2	&  $-$1 &	26 	&	25 	&	94.2 	&  &  1   &	17 	&	17 	& 95.8\\
         &     &     & GMMAI4	&	2 	&	14 	&	14 	&	93.0 	&  &  0   &	14 	&	14 	& 94.8\\
         &     &AR(1)& QIF	    &   0 	&	25 	&	25 	&	94.4 	&  &  2   &	39 	&	41 	& 96.0\\
         &     &     &GMMAI2	&   0 	&	25 	&	25 	&	94.0 	&  &  1   &	17 	&	17 	& 95.8\\
         &     &     &GMMAI4	&   2 	&	14 	&	14 	&	95.4 	&  &  1   &	14 	&	14 	& 94.8\\
\hline
    0.5  & 200 &CS   & QIF	    &	0 	&	34 	&	33 	&	94.0 	&  &  1   &	82 	&	80 	& 94.0\\
         &     &     &GMMAI2	&	0 	&	35 	&	33 	&	93.4 	&  &  0   &	28 	&	27 	& 94.0\\
         &     &     &GMMAI4	&	4 	&	22 	&	20 	&	91.8 	&  & $-$1 &	22 	&	21 	& 92.6\\
         &     &AR(1)&QIF       & $-$2 	&	37 	&	35 	&	94.2 	&  & $-$3 &	80 	&	77 	& 93.6\\
         &     &     &GMMAI2	& $-$1 	&	38 	&	34 	&	93.0 	&  & $-$2 &	28 	&	27 	& 92.4\\
         &     &     &GMMAI4	& 5 	&	23 	&	21 	&	93.0 	&  & $-$1 &	23 	&	21 	& 92.8\\
\\
         & 500 &CS   & QIF	    &  $-$1 &	20 	&	21 	&	96.4 	&  &  2   &	54 	&	51 	& 93.2\\
         &     &     & GMMAI2	&  $-$1 &	20 	&	21 	&	96.0 	&  &  0   &	17 	&	17 	& 94.6\\
         &     &     & GMMAI4	&	2 	&	13 	&	13 	&	95.0 	&  &  0   &	14 	&	14 	& 94.2\\
         &     &AR(1)& QIF	    &  $-$1 &	22 	&	22 	&	95.2 	&  & $-$3 &	51 	&	49 	& 93.8\\
         &     &     &GMMAI2    &  $-$1 &	23 	&	22 	&	94.4 	&  & $-$2 &	18 	&	17 	& 94.0\\
         &     &     &GMMAI4	&    1 	&	13 	&	13 	&	96.0 	&  & $-$1 &	14 	&	14 	& 94.8\\
\hline
   0.8   & 200 & CS  & QIF	    &	0 	&	23 	&	22 	&	94.8 	&  & $-$1 &	98 	&	92 	& 93.3\\
         &     &     &GMMAI2	&	0 	&	23 	&	22 	&	93.7 	&  & 0 	  &	29 	&	27 	& 93.9\\
         &     &     &GMMAI4	&	2 	&	18 	&	17 	&	93.5 	&  & 0 	  &	24 	&	22 	& 92.7\\
         &     &AR(1)& QIF	    &   1 	&	25 	&	24 	&	95.8 	&  & 2 	  &	96 	&	90 	& 92.2\\
         &     &     &GMMAI2	&   1 	&	26 	&	24 	&	93.8 	&  & 0 	  &	28 	&	27 	& 93.8\\
         &     &     &GMMAI4	&   4 	&	19 	&	18 	&	93.4 	&  & 0 	  &	24 	&	22 	& 91.8\\
\\
         & 500 & CS  & QIF	    &	0 	&	14 	&	14 	&	95.8 	&  & $-$4 &	58 	&	58 	& 94.8\\
         &     &     &GMMAI2	&	0 	&	14 	&	14 	&	95.2 	&  & $-$2 &	17 	&	17 	& 93.8\\
         &     &     &GMMAI4	&	1 	&	11 	&	11 	&	93.2 	&  & $-$1 &	14 	&	14 	& 93.6\\
         &     &AR(1)& QIF	    &   0 	&	15 	&	15 	&	95.4 	&  &  1   &	58 	&	57 	& 94.4\\
         &     &     &GMMAI2	&   0 	&	15 	&	15 	&	95.8 	&  & $-$1 &	18 	&	17 	& 94.6\\
         &     &     &GMMAI4	&   1 	&	12 	&	11 	&	94.4 	&  & $-$1 &	14 	&	14 	& 95.2\\
\hline
\end{tabular}
}
\end{center}
{\footnotesize Note: See Table\ref{Tab-1}.}
\end{table}

\begin{table}[htbp]
\begin{center}
\caption{Simulation results of model (\ref{mod}) under different $\rho_X$ with sample size $n=300$.}\label{Tab-3}
{
\setlength{\tabcolsep}{1.0mm}
\begin{tabular}{lll ccccccccccc}
\hline
  & & & \multicolumn{5}{c}{$\beta_1$}&  &\multicolumn{5}{c}{$\beta_2$}  \\
\cline{4-8}\cline{10-14}
$\rho_X$& $\bSigma_X^1$ & method & Bias & SD & SE & CP($\%$) & RE &  & Bias & SD & SE & CP($\%$)& RE\\
\hline
  0.2 & CS  &  QIF   &	 0 	 &	28 	& 27  &	95.4  &	--    &  & $-$1  &	68 	& 66  &	94.0  &	--\\
      &     &GMMAI2  &	 0 	 &	29 	& 27  &	92.4  &	0.92  &  &  0 	 &	23 	& 22  &	94.4  &	8.61\\
      &     &GMMAI4  &   3 	 &	18 	& 17  &	92.2  &	2.47  &  & $-$1  &	19 	& 18  &	93.8  &	13.26\\
\\
      &AR(1)& QIF	 &	2 	 &	29 	& 29  &	95.0  & -- 	  &  & $-$3  &	67 	& 66  &	93.2  & --\\	
      &     &GMMAI2	 &	1 	 &	29 	& 28  &	93.8  &	0.98  &  &	0 	 &	26 	& 25  &	93.4  &	6.97\\
      &     &GMMAI4	 &	3 	 &	18 	& 17  &	92.2  &	2.48  &  &	0 	 &	20 	& 19  &	94.2  &	11.39\\
\hline
  0.5 & CS  & QIF    &	3 	 &	35 	& 33  &	93.0  &	--	  &  & $-$3  &	68 	& 66  &	93.4  &	--\\
      &     & GMMAI2 &  3 	 &	35 	& 33  &	93.0  &	0.97  &  &	2 	 &	30 	& 30  &	96.8  &	5.16\\
      &     & GMMAI4 &  3 	 &	19 	& 18  &	93.0  &	3.27  &  &	1 	 &	21 	& 21  &	95.4  &	10.22\\
\\
      &AR(1)& QIF	 &	$-$1 &	34 	& 32  &	94.0  &--	  &  &	2 	 &	69 	& 66  &	94.0  & --\\
      &     & GMMAI2 &	$-$2 &	34 	& 31  &	93.4  &	0.96  &  &	1 	 &	30 	& 28  &	93.6  &	5.45\\
      &     & GMMAI4 &	3 	 & 	19 	& 18  &	92.8  &	3.20  &  &	2 	 &	22 	& 21  &	92.8  &	9.67\\
\hline
  0.8 & CS  & QIF    &  1 	 &	39 	& 39  &	95.4  &	--	  &  & $-$9  &	70 	& 66  &	93.2  &	--\\
      &     & GMMAI2 &  0 	 &	40 	& 39  &	94.6  &	0.96  &  & $-$4  &	33 	& 33  &	94.8  &	4.57\\
      &     & GMMAI4 &  4 	 &	20 	& 19  &	93.4  &	3.74  &  & $-$2  &	23 	& 22  &	93.8  &	9.43\\
\\
      &AR(1)& QIF	 &	0 	 &	39 	& 38  &	94.6  &	 --   &  & $-$1  &	69 	& 66  &	94.6  & --\\
      &     & GMMAI2 &	$-$1 &	39 	& 37  &	93.8  &	0.96  &  & $-$2  &	33 	& 32  &	94.6  &	4.33\\
      &     & GMMAI4 &	4 	 &	20 	& 19  &	93.0  &	3.77  &  & $-$1  &	23 	& 22  &	93.4  &	8.62\\
\hline
\end{tabular}
}
\end{center}
{\footnotesize Note: $\bSigma_X^1$ represents the covariance matrix of $\bX_1$ and $\rho_X$ is the correlation coefficient; Bias is the mean bias, SD is the standard deviation, SE is the standard error, and CP is the coverage probability, all are based on $500$ replications; QIF represents Qu et al. (2000) estimator, GMMAI2 and GMMAI4 represent our GMM estimator with auxiliary information $\bphi_1^*-\bphi_2^*$ and $\bphi_1-\bphi_4$ respectively.}
\end{table}

\begin{table}[htbp]
\begin{center}
\caption{Hypothesis test results by different methods under model (\ref{mod}) when sample size $n=300$ and $\bSigma_Y$ have the CS structure with $\rho_Y=0.5$.}\label{Tab-4}
{
\setlength{\tabcolsep}{1.2mm}
\begin{tabular}{cccccccccccccccc}
\hline
\multicolumn{7}{c} {$H_{01}:\beta_1=\beta_1^0$ }  &  &  & \multicolumn{7}{c} {$H_{02}:\beta_2=\beta_2^0$ } \\
\cline{1-7} \cline{10-16}
$\beta_1^0$ &  & QIF  &  & GMMAI2 &  & GMMAI4   &  &  &  $\beta_2^0$ &   & QIF &  &   GMMAI2 &  & GMMAI4 \\
\hline
 0.50   &   & 0.0380&   &0.0443 &   &0.0464 &  &  & $-$0.50 &   & 0.0593&   &0.0403 &   &0.0487\\
 0.55   &   & 0.2929&   &0.3091 &   &0.6081 &  &  & $-$0.55 &   & 0.1033&   &0.3695 &   &0.5637\\
 0.60   &   & 0.8337&   &0.8277 &   &0.9980 &  &  & $-$0.60 &   & 0.3279&   &0.9002 &   &0.9959\\
\hline
\end{tabular}
}
\end{center}
{\footnotesize Note: QIF represents the method proposed by Qu et al.(2000), GMMAI2 and GMMAI4 represent the proposed test method with auxiliary information $\bphi_1^*-\bphi_2^*$ and $\bphi_1-\bphi_4$ respectively.}
\end{table}

\newpage

\begin{table}
\begin{center}
\caption{Point estimates (PE) and their standard errors (SE) for the real data study obtained from a sampled subset with sample size 1000.}\label{Tab-5}
{ 
\setlength{\tabcolsep}{0.9mm}
\begin{tabular}{cccccccccccc}
\hline
  & \multicolumn{3}{c}{$\beta_1$}&  &\multicolumn{3}{c}{$\beta_2$}& &\multicolumn{3}{c}{$\beta_3$}\\
\cline{2-4}\cline{6-8} \cline{10-12}\\
   Method        & PE       & SE       & p-value   &   &PE        &   SE        & p-value  &   &   PE       &   SE  & p-value\\
\hline
  QIF         & $-$0.1209  &  0.0187   & $<$0.001   &   & 0.4186     & 0.0178  & $<$0.001&   &0.3971 & 0.0165  & $<$0.001\\
  GMMAI2(I)   & $-$0.1292  &  0.0162   & $<$0.001   &   & 0.4174     & 0.0177  & $<$0.001&   &0.4043 & 0.0164  & $<$0.001\\
  GMMAI4(I)   & $-$0.1240  &  0.0143   & $<$0.001   &   & 0.4617     & 0.0178  & $<$0.001&   &0.3983 & 0.0155  & $<$0.001\\
\\
  GMMAI2(II)  & $-$0.0939  &  0.0160   & $<$0.001   &   & 0.4497     & 0.0177  & $<$0.001&   &0.3996 &  0.0155 & $<$0.001 \\
  GMMAI4(II)  & $-$0.0907  &  0.0143   & $<$0.001   &   & 0.4865     & 0.0143  & $<$0.001&   &0.3860 & 0.0132  & $<$0.001\\
\\
  GMMAI2(III) & $-$0.1053  &  0.0160   & $<$0.001   &   & 0.4451     & 0.0167  & $<$0.001&   &0.4125 & 0.0164  & $<$0.001\\
  GMMAI4(III) & $-$0.1047  &  0.0153   & $<$0.001   &   & 0.4363     & 0.0157  & $<$0.001&   &0.4474 & 0.0161  & $<$0.001\\
 \hline
\end{tabular}
}
\end{center}
{\footnotesize Note: ``I" represents grouping the subjects by gender and math ability score in Grade $3$, ``II" stands for grouping the subjects by the math and reading ability score in Grade $3$ and ``III" indicates that we dividing the subjects into groups by the reading ability scores in Grade $3$ and $8$. }
\end{table}

\end{document}